\documentclass[%
superscriptaddress,
preprint,
 amsmath,amssymb,
 aps,
pre,
]{revtex4-1}

\usepackage{graphicx}
\usepackage{dcolumn}
\usepackage{bm}
\usepackage[colorlinks=true, linkcolor=blue, citecolor=blue]{hyperref}

\usepackage{amsmath}

\begin{document}


\title{Generation of gravity waves by pedal-wavemakers}

\author{\textcopyright Isis Vivanco}
\affiliation{Departamento de F\'isica, Universidad de Santiago de Chile, Av. Ecuador 3493, Estaci\'on Central, Santiago, Chile.}

\author{Bruce Cartwright}
\affiliation{
Pacific Engineering Systems International, 277\textendash 279
Broadway, Glebe, New South Wales 2037, Australia}
\affiliation{The University of Newcastle, Callaghan, New South Wales 2308,
Australia
}

\author{A. Ledesma Araujo}
\affiliation{Departamento de F\'isica, Universidad de Santiago de Chile,
Av. Ecuador 3493, Estaci\'on Central, Santiago, Chile.}

\author{Leonardo Gordillo}
\affiliation{Departamento de F\'isica, Universidad de Santiago de Chile, Av. Ecuador 3493, Estaci\'on Central, Santiago, Chile.}

\author{Juan F. Mar\'in}
\email{juan.marin.m@usach.cl}
\affiliation{Departamento de F\'isica, Universidad de Santiago de Chile,
Av. Ecuador 3493, Estaci\'on Central, Santiago, Chile.}

\date{\today}

\begin{abstract}
Experimental wave generation in channels is usually achieved through wavemakers (moving paddles) acting on the surface of the water. Although practical for engineering purposes, wavemakers have issues: they perform poorly in the generation of long waves and create evanescent waves in their vicinity. In this article, we introduce a framework for wave generation through the action of an underwater multipoint mechanism: the pedal-wavemaking method. Our multipoint action makes each point of the bottom move with a prescribed pedalling-like motion. We analyse the linear response of waves in a uniform channel in terms of the wavelength of the bottom action. The framework naturally solves the problem of the performance for long waves and replaces evanescent waves by thin boundary layer at the bottom of the channel. We also show that a proper synchronisation of orbital motion on the bottom can produce waves that mimic deep water waves. This last feature has been proved to be useful to study fluid-structure interaction in simulations based on smoothed particle hydrodynamics.
\end{abstract}

\maketitle


\section{Introduction\label{Sec:Introduction}}

 The engineering of surface gravity waves in water channels is a challenging problem in both numerical and experimental setups. The importance of a controlled and efficient generation of waves with prescribed amplitude and phase lies not only in the fundamental research on wave propagation and wave instabilities \cite{Cross2009,Lighthill1978} but also in practical applications in hydraulics. For instance, in the recreation of tsunami-wave dynamics in laboratory-scale systems, the employed wave-generation technique plays an important role in the resulting tsunami waveform and dynamics \cite{Grilli2013,Jamin2015}. Thus, implementing an efficient source model that is also a realistic representation of a natural wave-source is a relevant issue. Another example of the importance of realistic and efficient wave-sources comes from the field of fluid-structure interaction \cite{Cunningham2014,Jamin2015}, in experimental and numerical tests of ships and structures interacting with regular waves in the sea \cite{Cartwright2010}. The characterisation of the mechanical fatigue in the hull of a ship subjected to multiple collisions with waves is fundamental to design norms and to develop new technologies in shipbuilding \cite{Cartwright2012,Lloyd1991}.

Hybridised mesh-free and finite methods are widely used as a numerical tool for the prediction of the structural response of floating structures to waves \cite{Cartwright2012,Cartwright2010,Cunningham2014}. Among the Lagrangian mesh-free particle methods, Smoothed Particle Hydrodynamics (SPH) has gained great interest \cite{Li2002,Liu2003,Sigalotti2003,Liu2010,Cartwright2010,Cartwright2012,Wang2016}. In such a method, the absence of a discretisation mesh allows to perform otherwise complex and numerically expensive calculations that naturally occur in systems with high deformations, such as shock waves \cite{Sigalotti2009,Sigalotti2006,Marrone2011}, high vorticity flows \cite{Sun2018}, jets \cite{dePadova2020}, drops \cite{Melean2004,Melean2005}, bubbles \cite{Ming2017}, and spraying \cite{Gnanasekaran2019}. The fluid phase is discretised into a distribution of small fluid particles, whose dynamics is influenced by neighbouring particles that lie within a support domain. For a given particle, the effect of the surrounding fluid is determined by a weighted sum over all the remaining particles in the system. Although the SPH method has some drawbacks and numerical issues, such as the well-known tensile instability at low Reynolds numbers \cite{Melean2004,Sigalotti2008}, the method can handle splash and violent free surface events. For this reason, the SPH method has become widely used in numerical studies of the interactions of ships and structures with sea waves. 

To generate surface waves in water, the usual technique is the inclusion of a \emph{wavemaker}\textendash{} an oscillating paddle attached to a wall at one of the boundaries of the fluid domain. Such a technique has proven to be very efficient to generate short-wavelength surface waves and has been included in mesh-free particles methods to obtain realistic simulations of floating objects interacting with a train of severe waves \cite{Cartwright2010}. Wavemakers can also be used to cancel incident waves \textendash{} the so-called \emph{water-wave active absorbers} \cite{Milgram1970,Schaffer2000}\textendash , and have shown better performance than passive methods such as beaches or meta-materials \cite{Ouellet1986}. However, the use of wavemakers has side effects that may represent a drawback in some situations. According to the fundamental theory of wavemakers \cite{Havelock1929}, the oscillations of the wall generates two kinds of waves: a short-range evanescent wave and a long-range radiative wave. Therefore, to provide full control of the amplitude and phase of the generated waves, the region of study must be placed far enough from the wavemaker to avoid non-desired effects from the evanescent wave. Numerical simulations and experiments using this strategy require a large fluid domain, which is costly in numerical operations and experimental resources. Indeed, many experimental studies of tsunamis and rogue waves required large water tanks with huge paddles and parts (see Refs. \cite{Bathuy2009,Briggs1995,Dematteis2019,Mcallister2019}
for some examples). A way to circumvent this issue in numerical setups is to solve the hydrodynamic equations in regions with different resolutions: A low-resolution fluid domain where waves are generated by the paddles and travel through a long distance \textendash so that evanescent waves decay\textendash , and a high-resolution domain \textendash{} or zone of interest \textendash where the structure is placed in interactions with the incident waves. However, this strategy carries along with
another drawback, which is the energy dissipation intrinsic to the SPH method. A long low-resolution domain also provides dissipation of the radiative wave, thus requiring high-energy of oscillations at the wave paddle. Moreover, paddle wavemakers present another drawback in some numerical and experimental situations, which is their poor performance in the generation of long-wavelength waves: The paddle must oscillate with very high amplitude to produce a long-wavelength wave with a relatively small amplitude \cite{Ursell1960}.

In this article, we propose a new technique for the generation of surface waves in a water channel: the \emph{pedal-wavemaking method}. The method consists in moving the floor of the channel in a pedalling-like elliptical motion prescribed by the Airy inviscid theory of deep-water waves. Through this technique, we can design waves that emulate deep gravity waves. We demonstrate numerically and theoretically that the technique can generate long-wavelength surface-waves, which are elusive for systems with wavemakers. Moreover, since the wave generation technique does not produce an evanescent surface wave, the technique does not require a large fluid domain to produce controlled waves.

The  article is organised as follows. Section \ref{Sec:Methods} gives a summary of the wave-generation techniques that are relevant in our study: by paddle-wavemakers and pedal-wavemakers. We also give the mathematical model to study wave generation phenomena in a finite-depth tank with a pedalling-like moving bottom. Section \ref{Sec:Results} shows the results from SPH simulations of regular wave generation. We provide a qualitative comparison between both methods and the Airy theory of deep water waves. We also show in this section the results from the theoretical model of pedal wavemakers. We discuss our results in section \ref{subsec:Discussion}. Conclusions and final remarks are given in Section \ref{sec:Conclusions}.

\section{Methods \label{Sec:Methods}}



\subsection{Wave generation using a hinged-paddle with  SPH simulations\label{subsec:PaddleMethod}}

\begin{figure}
\centering{}\includegraphics[width=1\linewidth]{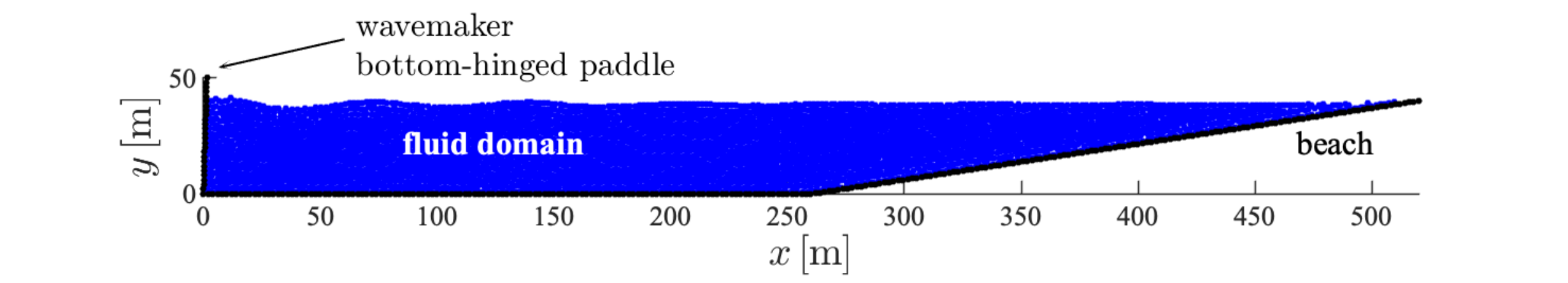} \caption{Numerical setup for the tow-tank model. Waves are generated by a hinged-paddle wavemaker placed at the left boundary of a two-dimensional fluid layer of depth $h=40\,${[}m{]} and length $L=520\,[\mbox{m]}$. A beach is placed at the right boundary to reduce reflections.
\label{fig:01}}
\end{figure}

The first method of wave generation considered in this article is the well-known hinged-paddle technique. Figure~\ref{fig:01} shows the tow-tank model, so-called as it is of the form used to measure the response of a model ship being towed through the waves. The tow-tank has a typical configuration of a hinged-paddle wavemaker for generating waves, and a beach for the dissipation of the waves \cite{Ozbulut2020}. A two-dimensional layer of water is contained in a numerical tank of length $L=520\,[\mbox{m]}$ and depth $h=40\,[\mbox{m]}$. A rigid floor-hinged paddle wavemaker is placed at one end and a gently sloping beach is placed at the other end to reduce reflections. We choose the wavemaker amplitude and frequency to give a wave with wavelength $\lambda=60\,[\mbox{m]}$ and amplitud $\eta=2\,\mbox{[m]}$ close to the paddle.

We performed numerical simulations of the hydrodynamic equations using the \emph{smoothed particle hydrodynamics} (SPH) method \cite{Liu2003,Liu2010,Wang2016}. In the SPH method, the field variables and gradients are obtained in a Lagrangian framework through an interpolating kernel\textendash{} the so-called\emph{ kernel approximation} \textendash that gives a smooth estimate of the physical properties of the fluid, such as the density, pressure, and velocity. The formal space-discretisation of hydrodynamic equations is obtained through a \emph{particle approximation,} that estimates smoothed integrals as a sum on a set of fluid particles. Finally, time integration is achieved using standard time-integration methods, such as Runge-Kutta schemes and predictor-corrector methods. We have used $N=4785$ fluid particles to simulate the system of Fig.~\ref{fig:01}. The sea-bottom, the sloping beach and the hinged paddle are shell elements with a prescribed velocity value: time-dependent for the latter and fixed for the earliest.  The shell elements and SPH particles interact through industry-common contact-interface algorithms.

In Section \ref{Sec:SimulationsWavemakers} we show that the waves generated by paddles has poor efficiency generating long waves and creates evanescent waves on its vicinity. To circumvent these issues, we have developed a new wave generation technique described in Section~\ref{subsec:PedaTechnique}.

\subsection{The pedal-wavemaking technique\label{subsec:PedaTechnique}}

\begin{figure}
\includegraphics[scale=0.8]{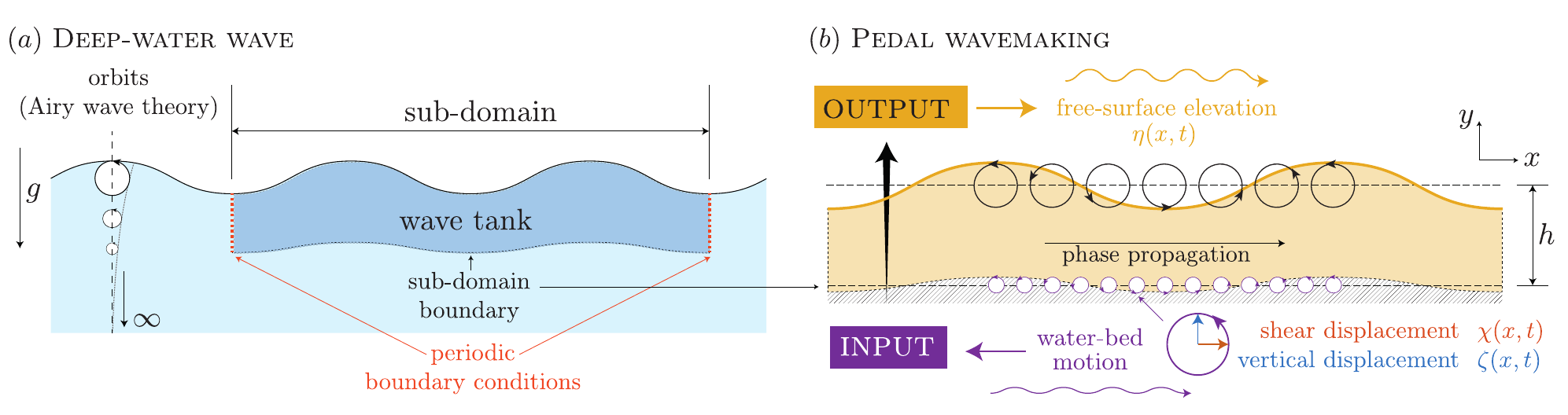}\caption{(a) Schematisation of the sub-domain (wave tank) in infinitely deep water. The boundary motion at the bottom of the wave tank is prescribed by the Airy wave inviscid theory. (b) Illustration of the pedal-wavemaking technique. The water bed motion is given by a pedalling-like motion, with horizontal and vertical displacements prescribed by the Airy wave theory at depth $y=-h$. This motion of the water bed represents gives rise to a travelling wave in the free surface. \label{fig:02}}
\end{figure}

Waves in fluids are naturally generated by pressure perturbations applied either in the free surface of the water or in the bottom layer. Although such techniques of wave generation are not commonly used in experimental situations, the underlying mechanism occurs naturally in the sea. For instance, waves in the sea are mainly a consequence of wind and atmospheric forcing. Such forcing makes surface fluid-particles to describe periodic\textendash{} circular or elliptical \textendash trajectories \cite{Stoker2011}, like those depicted in Fig. \ref{fig:02}(a). The collective motion of fluid particles at the surface is part of the generated travelling surface-wave. For deep-water waves, the response of fluid particles  below the surface is given by the Airy linear-wave theory \cite{Lamb1932}. Particles in the bulk follow the path of circular orbits whose radius decreases exponentially as we go deeper into the fluid.

To develop a new alternative technique for wave generation, we wondered on the inverse problem: can we generate gravity waves in the free surface imposing small circular orbits of fluid particles in the bulk? Instead of considering an infinitely deep-water wave system, we consider a reduced subdomain \textendash{} the wave tank \textendash{} shown in Fig. \ref{fig:02}(a). The floor of the subdomain is considered as a boundary with some prescribed motion. Using periodic boundary conditions at the left and right boundaries of the wave tank, orbits in infinitely deep water channels can be recreated in this small sub-domain using moving boundary conditions at the bottom, namely, a \emph{pedal-wavemaking motion}. 

Figure \ref{fig:02}(b) illustrates our pedal-wavemaking technique in a wave tank of depth $h$. In the SPH formulation, we have divided the bottom boundary into small shell elements with prescribed individual motions according to the Airy wave conditions at the location of each shell. This motion of individual elements can be regarded as an underwater multi-point mechanism that introduces energy to the fluid. The boundary conditions assigned to these moving shell elements at $y=-h$ are elliptical orbits, which can be decomposed into horizontal and vertical displacements $\chi(x,t)$ and $\zeta(x,t)$, respectively. Thus, the elliptical trajectories of the particles at the bottom of the wave tank are introduced by moving the bottom itself with a given orbital trajectory, resembling a pedalling-like motion.

As we will see in Section \ref{subsec:GravityWaves}, the pedalling wavemakers efficiently generate long-waves. Notice that the generation of gravity waves by pedal-wavemakers, i.e. through the bed motion, is similar to the tsunami generation, which are long waves generated by the sudden uplift of the marine base during earthquakes \cite{Kajiura1963}. Tsunamis lie in the long-wavelength spectrum of the oceanic waves \cite{Munk1950}. Pedal-wavemaking technique is thus the natural way to generate waves in the region of long wavelengths of the  spectrum.

\subsection{Theoretical model\label{Sec:Theory} }

A full theoretical approach into the pedal-wavemaking technique considering all the features of a Newtonian fluid and the right boundary conditions is required. We regard the two-dimensional system of Fig~\ref{fig:02}(b) as an infinitely long channel filled with an incompressible viscous liquid up to a uniform depth $h$. The equations that describe the fluid are given by the Navier-Stokes and incompressibility equation,\begin{subequations}\label{eq:Ns-incompress}
\begin{equation}
\partial_{t}\mathbf{u}+(\mathbf{u}\cdot\boldsymbol{\nabla})\mathbf{u}=-\frac{1}{\rho}\boldsymbol{\nabla}P+\nu\nabla^{2}\mathbf{u}-g\mathbf{\hat{y}}\label{eq:NS}
\end{equation}
\begin{equation}
\boldsymbol{\nabla}\cdot\mathbf{u}=0,\label{eq:incompress}
\end{equation}
\label{Eq:FluidEquations}
\end{subequations}
where $\mathbf{u}=(u,v)$ is the velocity field, $P$ is the pressure, $g$ is the acceleration of gravity, $\nu$ is the kinematic viscosity, and $\rho$ is the density of the fluid. We decompose the water-bed motion into a vertical and horizontal periodic displacement $\zeta(x,t)$ and $\chi(x,t)$, respectively. A pedalling-like motion of points in the water-bed with angular frequency $\omega$ can be described giving an appropriate phase difference,
\begin{equation}
\left(\chi,\zeta\right)=\Re\left(ix_{b}e^{i\left(kx-\omega t\right)},y_{b}e^{i\left(kx-\omega t\right)}\right),\label{eq:input}
\end{equation}
where $x_{b}$ and $y_{b}$ are complex amplitudes. The collective motion of points in the bottom generates a travelling-wave-like motion of the water bed, i.e. a phase propagation in the positive $x$-direction with wavenumber $k$. The output in the free-surface elevation is a gravity-wave response with amplitude $\eta$.

\section{Results \label{Sec:Results}}


\subsection{SPH simulations of regular wave generation by hinged paddles \label{Sec:SimulationsWavemakers}}

\begin{figure}
\centering{}\includegraphics[width=.9\linewidth]{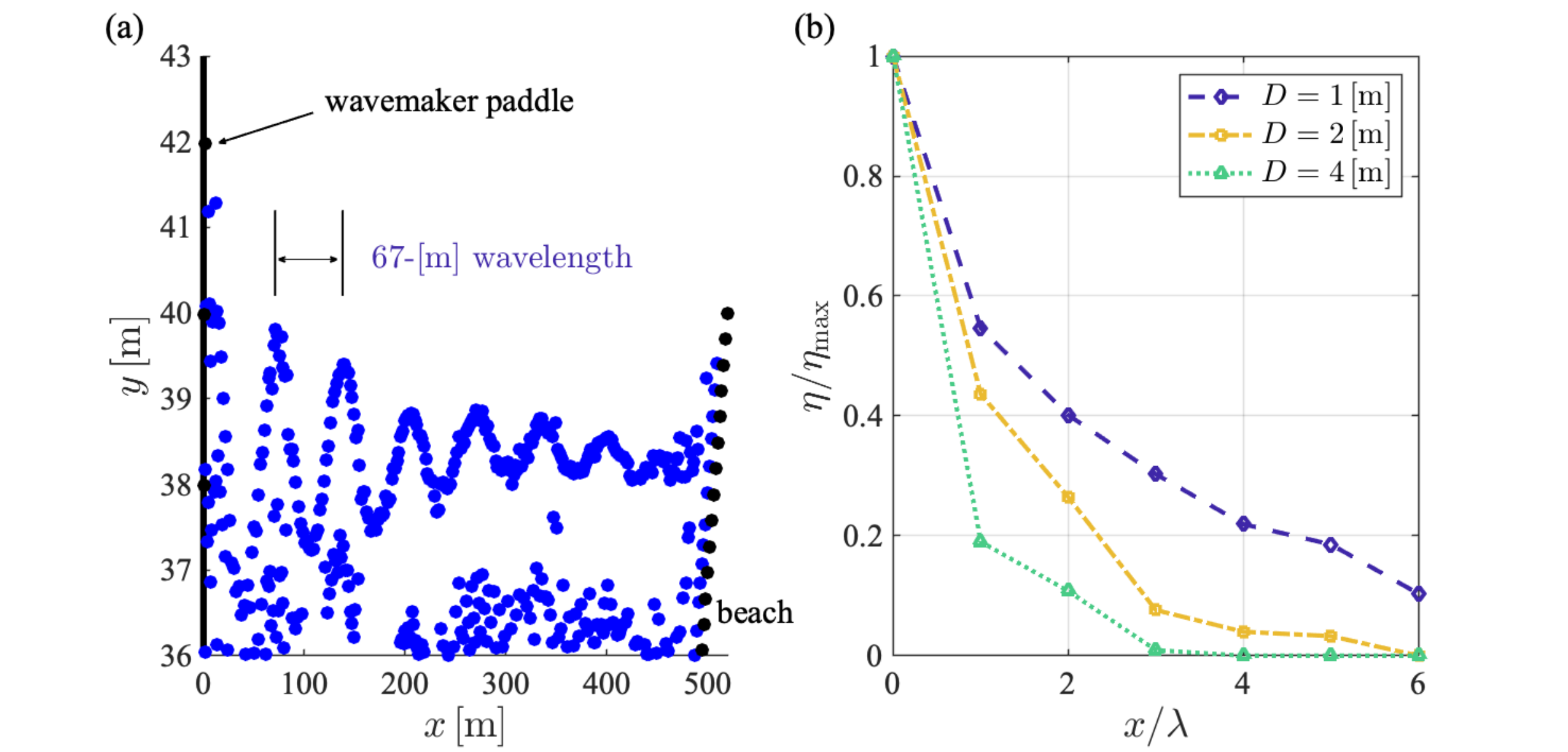} \caption{Results from SPH simulations of the tow-tank model. (a) Fluid particles in the SPH simulations, showing that the wave amplitude from the full-size wave tank decreases with distance from the wavemaker. (b) Normalised wave amplitude as a function of the distance from the wavemaker for $\lambda=60\,[\mbox{m]}$ and fluid particles with the indicated diameters. Waves generated with small-diameter particles exhibit less energy dissipation. The simulation was run for $120\,\mbox{[s]}$, which corresponds to approximately $20$ periods of the wavelength. 
\label{fig:03}}
\end{figure}

Our SPH numerical simulations using the well-known method of hinged-paddles for  wave generation (see Section \ref{subsec:PaddleMethod}) employed a tow-tank model shown in Fig.~\ref{fig:01}. Our results show that the wave paddle generates a travelling surface-wave with wavelength $\lambda\simeq67\,[\hbox{m}]$, as depicted in Fig.~\ref{fig:03}(a). However, as the wave propagates towards the beach, its amplitude is significantly reduced due to viscosity. From Fig.~\ref{fig:03}(a) we notice that near the hinged-paddle the wave is approximately $2\,\mbox{[m]}$-amplitude, whereas near the beach is below a half-meter amplitude. This observation is confirmed in Fig.~\ref{fig:03}(b), where we show the normalised wave amplitude as a function of the distance from the wavemaker. It is important to remark that the observed decay is due not only to physical viscosity of the fluid but also to nonphysical energy losses at each of the many numerical smoothing calculations performed as the wave propagates
through the tank \cite{McCue2006}. Since the object under study in interaction with waves must be placed away from the paddle to avoid the disturbances from the evanescent waves, such decay in the wave amplitude becomes a problem.

McCue and co-workers \cite{McCue2006} have shown in similar numerical setups that smaller SPH particles provide less energy dissipation. Following their ideas, we used fluid particles with different diameters $D$. In Fig.\ref{fig:03}(b) we show the normalised wave amplitude obtained for three diameter values of SPH particles. Indeed, less energy dissipation is observed for the smallest particles, agreeing with the findings of McCue and coworkers \cite{McCue2006}. However, the results are still unacceptable for some applications requiring uniform wave amplitudes, such as in the study of ship motion in waves \cite{Cartwright2012}. To limit wave-amplitude loss to around $10\%$ over $6$ wavelengths would require $D\sim0.1\,[\mbox{m]}$ \cite{Cartwright2012}. The use of such small particles in a $260\,\mbox{[m]}$-length wave-tank would dramatically increase the computational cost. Other authors \cite{Jones2006,Guilcher2007} suggested an improved time-stepping algorithm as a possible solution to reducing the wave decay. De Padova and coworkers \cite{DePadova2009} also showed high dissipation in their generated waves in a tank with a sloping floor, including both regular and irregular waves. Such a configuration ultimately led to breaking on a shore, so even those waves were not directly comparable to a constant-depth wave tank for ship-motion predictions. De Padova and coworkers concluded that a small value of one of the empirical coefficients of the SPH artificial viscosity was required for numerical stability, but the result was still too dissipative for accurate wave reproduction.

Another well-known important drawback of the hinged-paddle wavemaker is its poor performance in the generation of long-wavelength waves. Let us define the efficiency of the wavemaker as
\begin{equation}
 \varepsilon:=\frac{a(h/\lambda)}{s},   
\end{equation}
with $s$ the displacement of the paddle and $a$ the amplitude of the radiative wave, which is a function of the ratio $h/\lambda$. Here, $h$ is the depth of the channel and $\lambda$ is the wavelength \cite{Ursell1960}. It can be shown that in the short-wavelength limit
\begin{equation}
    \lim_{h/\lambda\to\infty}\varepsilon= 1,
\end{equation}
which means that the amplitude of the wavemaker must be of the order of the wave amplitude, i. e. $s\sim a$. However, the efficiency decays as $h/\lambda$ in the long-wavelength limit $h/\lambda\to0$, which means that the wavemaker must oscillate at an amplitude of the order of the wavelength\emph{, }i. e. $s\sim a\lambda/h$. Thus, the paddle must oscillate with very high amplitude to produce a long-wavelength wave with a relatively small amplitude. Consequently, the generation of waves by replicating the wave tank physics using paddle-type wavemakers in the SPH formulation is not a viable approach to predict the response of structures to long waves.

\subsection{Generation of long waves by pedal wavemakers\label{subsec:GravityWaves}}

\begin{figure}
\includegraphics[scale=0.5]{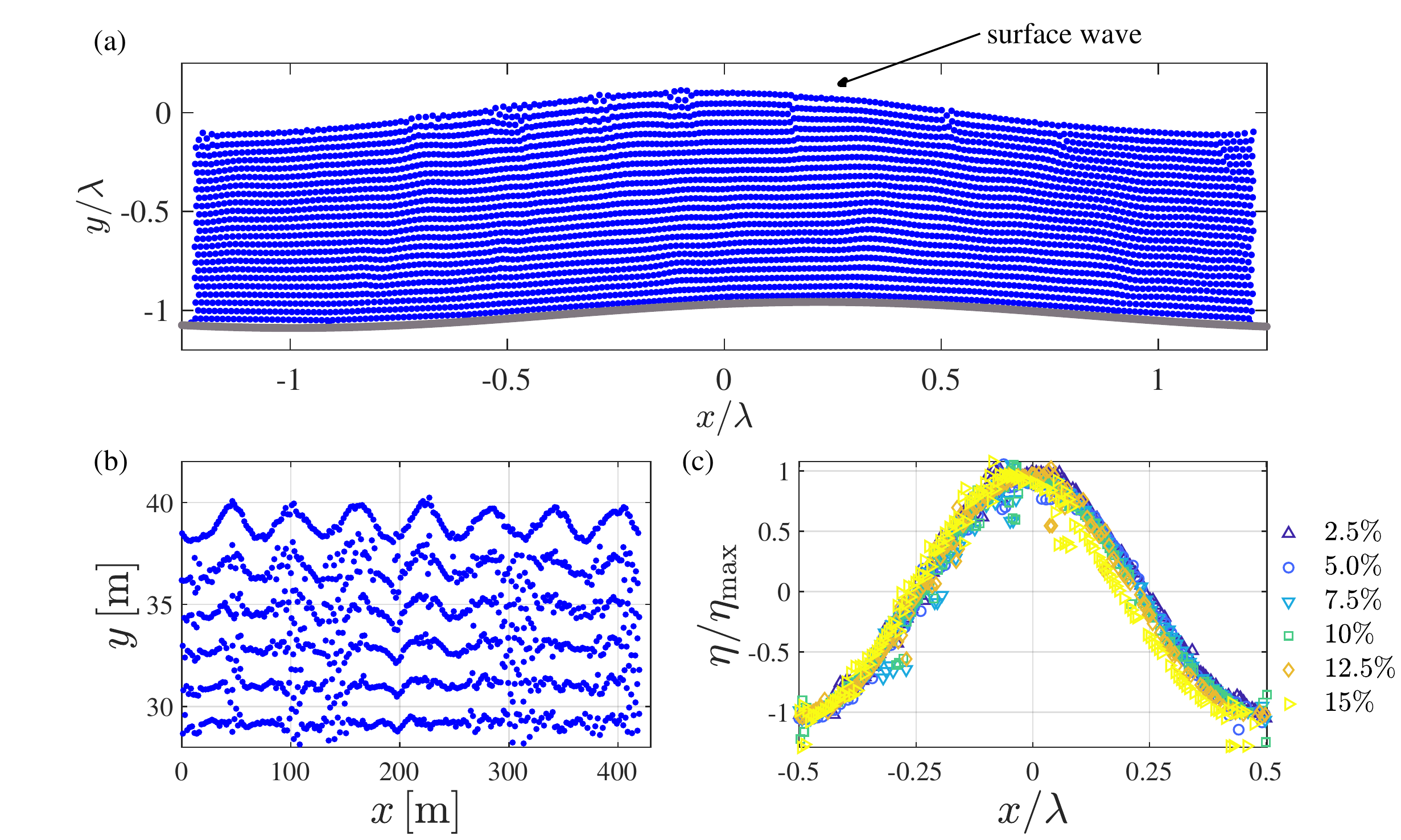}\caption{SPH simulations of long-wavelength generation using the pedal-wavemaking technique. (a) A typical fully developed long-wavelength wave. (b) Wave generated with $\lambda=60\,\mbox{[m]}$ and $f=0.8\,\mbox{[Hz]}$ by pedal-wavemaking. A pedalling-like motion of diameter $0.88\,[\mbox{m]}$ at the bottom generates a $2\,[\mbox{m]}$-amplitude surface wave. (c) Normalised profiles of surface waves obtained in wave-tanks of different depths \textendash{} see explanation in the text. The observed wave amplitude was $2\%$ of the wavelength. The roughness of the surface is due to the discretisation of the fluid into particles. Each profile has been normalised by its actual wave amplitude $\eta$.\label{fig:04}}
\end{figure}

The pedal-wavemaking technique introduced in Section \ref{subsec:PedaTechnique}, does not show any wave decay or evanescent waves near the source and is able to generate of long-wavelength waves as we will show in the following. Figure~\ref{fig:04} depicts typical long waves generated with the pedal-wavemaking method. In Fig.~\ref{fig:04}(a) we generated a wave with $\lambda=L_{wt}$, with $L_{wt}=2.5\,\mbox{[m]}$ the length of the wave-tank, and in Fig.~\ref{fig:04}(b) we produced a wave with $\lambda=60\,\mbox{[m]}$. In both cases the waves were obtained with the pedal-wavemaking technique using SPH particles with diameter $D=2\,[\mbox{m]}$. Notice that there is no loss of wave amplitude along the wave tank. However, it is important to remark that the pedal-wavemaking technique does not overcome the losses inherent to the SPH formulation. The pedal wavemaking relies on the interaction of a wave with the floor in shallow depths.

In the simulations of Fig.~\ref{fig:04}, the wave is developed simultaneously everywhere throughout the tank in response to the pedalling motion on the floor. Thus, fully developed waves rapidly fill the wave tank. Waves due to the pedalling motion are injected throughout the seabed and not at one extreme of the tank as the case for a single hinged-paddle. We observed that for some optimal values of the frequency and wavelength, a small pedalling motion in the seabed can generate a relatively high-amplitude gravity wave. Figure \ref{fig:04}(b) shows a $2\,[\mbox{m]}$-amplitude surface wave designed with $\lambda=60\,\mbox{[m]}$ and frequency $f=0.8\,\mbox{[Hz]}$. We used SPH particles with diameter $D=1${[}m{]}, and the floor of the wave tank was divided into eight equal segments with a $0.88\,[\mbox{m]}$-diameter pedalling-like motion. The amplitude of the generated wave is greater than the diameter of the pedalling by a factor greater than two. This observation suggests that the system is near some resonance condition at the given wavelength and frequency. Further studies characterising such resonance will be published elsewhere.

To characterise how the depth of the wave-tank affects the properties of waves generating through pedal-wavemaking, let us introduce the following dimensionless variables
\begin{equation}
    \tilde{\eta}\equiv\frac{\eta}{\lambda},\quad\quad
    \tilde{h}\equiv \frac{h}{\lambda},\quad\quad \tilde{D}\equiv\frac{D}{\lambda}.
\end{equation}
From the simulation shown in Fig. \ref{fig:04}(b) one obtains $\tilde{\eta}\simeq0.03$, which is in accordance with the long-wave regime. We perform SPH simulations using different depths in the range $0.025\leq\tilde{h}\leq0.15$. Normalised surface profiles at a fixed time are shown in Fig.~\ref{fig:04}(c) for the indicated values of the depth. For all the floor depths under study, we notice that the profile of the wave seems to be unchanged. This suggests that our wave-generation technique using pedal-wavemakers is robust to the depth of the wave-tank, and thus, it emulates deep-water waves under controlled situations.

\begin{figure}
\includegraphics[width=\linewidth]{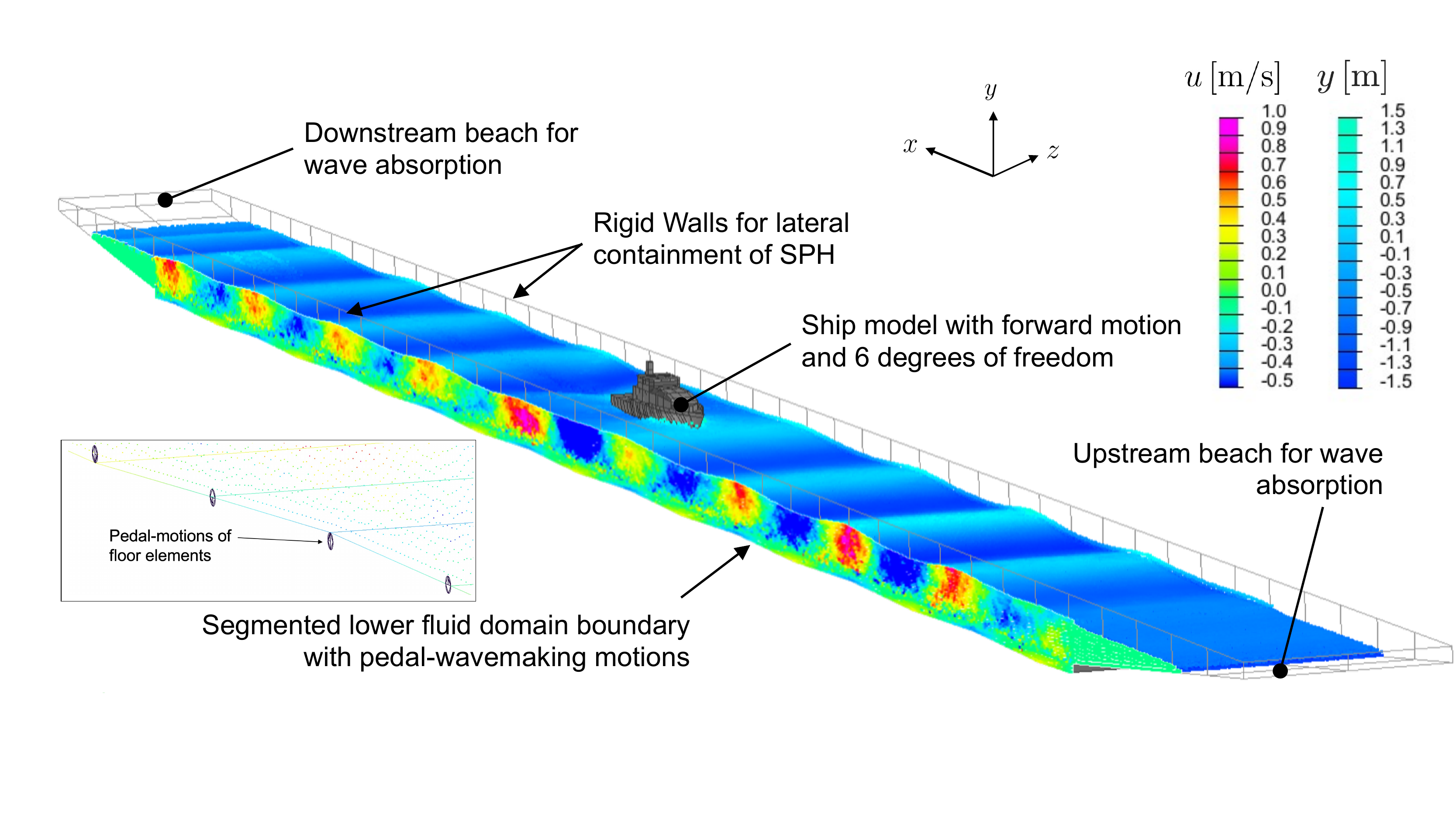}
\caption{Visual representation of a three-dimensional numerical model of a floating rigid ship interacting with waves. A $1.5[\hbox{m}]$-wave is generated and sustained using only the pedal-wave making floor motions.
\label{fig:05}}
\end{figure}

Pedal wavemakers can be used solely to generate and sustain travelling waves, as we show in the three-dimensional numerical model of Fig.~\ref{fig:05}. A ship model with appropriate degrees of freedom can be placed in interaction with waves in a wave tank. In Fig.~\ref{fig:05}, we generate a $1.5[\hbox{m}]$-height travelling wave using only four shell elements per wave for the water-bed. The nodes of such elements were prescribed with time-dependent boundary conditions to enable a moving-floor. Notice from Fig.~\ref{fig:05} that the maximum horizontal velocity is located in the bulk of the fluid, just below the surface. We will come back to this point in Section \ref{subsec:Emulating}.

\subsection{Emulating deep-water waves in a wave tank\label{subsec:Emulating}}

\begin{figure}
\includegraphics[width=1\linewidth]{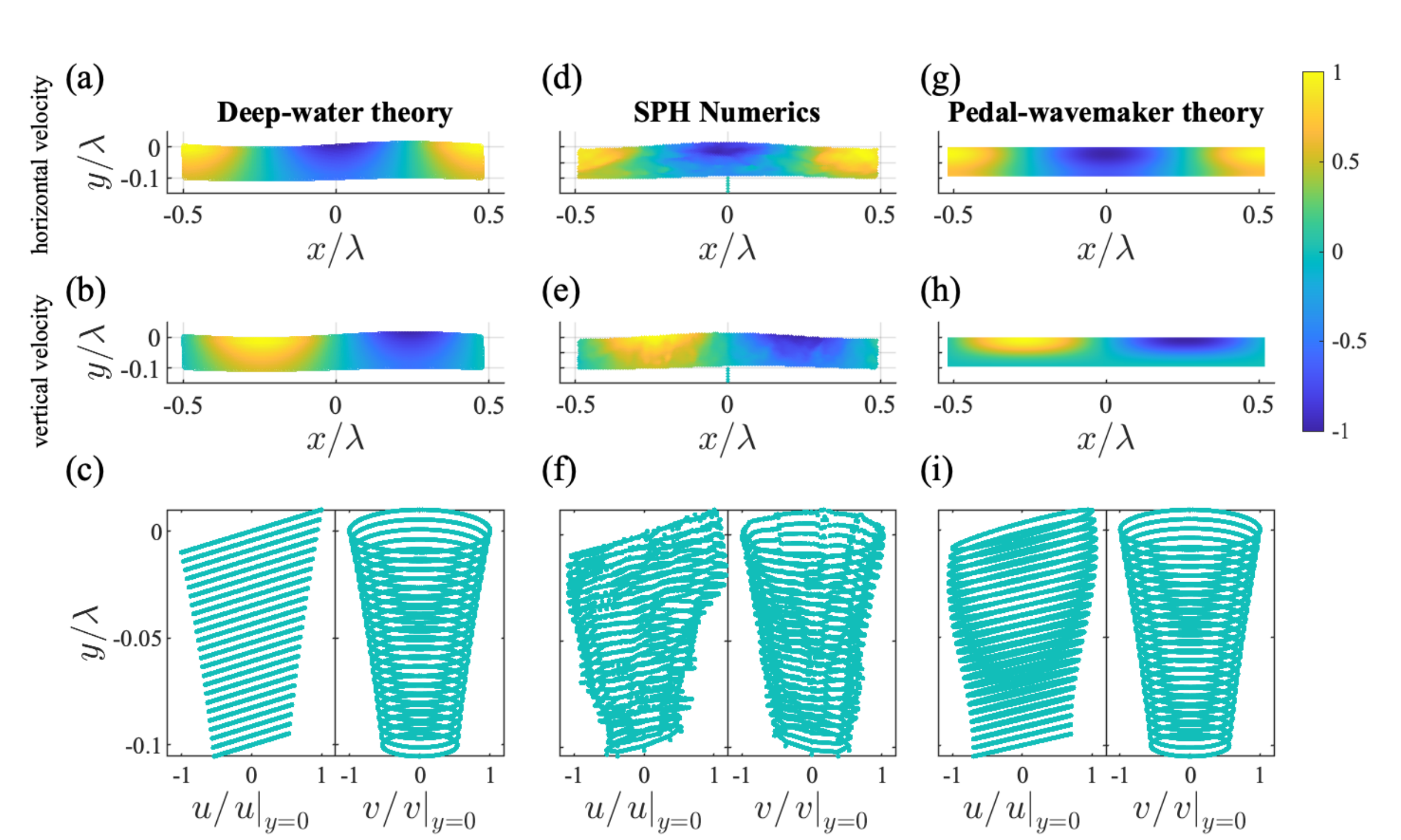}
\caption{Comparison between the Airy theory for inviscid deep water waves (left column), SPH numerical simulations using pedal-wavemakers at a depth of $10\%$ of the wavelength (central column), and the theory of pedal-wavemakers with $\bar\theta=10^{-1}$ and $\bar{{\lambda}}=10$ (right column). The diameter of the pedalling-like motion was $0.01\,h$. The upper row shows typical profiles of the horizontal velocity, whereas the central row shows the corresponding vertical velocity profiles. The colour scale indicates the normalized value of the speed. The normalised vertical and horizontal velocity profiles of fluid particles within the wave are depicted in the lower row. \label{fig:06}}
\end{figure}

Figure~\ref{fig:06} shows the through-depth velocity profiles in the horizontal and vertical directions for $\tilde{h}=0.10$. We have verified that numerical results are qualitatively similar for any wave-tank depth. The horizontal and vertical velocity profiles according to the Airy inviscid deep-water-wave theory are shown in Fig.~\ref{fig:06}(a)
and Fig.~\ref{fig:06}(b), respectively, for the given values of the parameters. According to the deep-water-wave theory, the amplitude of the wave decreases exponentially with the depth as $\exp(ky)$, where $y$ is the vertical coordinate and $k=2\pi/\lambda$. The horizontal and vertical components of the velocity can be easily obtained knowing the frequency of oscillations of the wave, namely $\omega=2\pi f.$ Let $u$ and $v$ be the horizontal and vertical components of the velocity of a fluid particle placed at an average depth $y_{p}$, respectively. Thus, according to the deep-water-wave theory, trajectories of fluid particles in phase spaces $\{(u,y)\}$ and $\{(v,y)\}$ follow\begin{subequations}
\begin{equation}
y=y_{p}+\left.\Delta y\right|_{y=0}\left(\frac{u}{\left.u\right|_{y=0}}\right),\label{ParticleTrajx}
\end{equation}
\begin{equation}
\left(\frac{y-y_{p}}{\left.\Delta y\right|_{y=0}}\right)^{2}+\left(\frac{v}{\left.v\right|_{y=0}}\right)^{2}=\exp\left(2ky\right),\label{ParticleTrajy}
\end{equation}
\end{subequations}
which corresponds to straight lines and ellipses
for the vertical and horizontal components of velocity, respectively, as depicted in Fig. \ref{fig:06}(c). From Eq. \eqref{ParticleTrajx}, it follows that the slope of $y(u)$ is positive and increases in absolute value with the wavelength. Likewise, from Eq. \eqref{ParticleTrajy}, it follows that the area enclosed by the elliptical trajectories decreases with the depth.

Figure \ref{fig:06}(d) and \ref{fig:06}(e) shows the horizontal and vertical velocity profiles obtained from numerical simulations in SPH using the pedal-wavemaking technique. The similarities with the corresponding profiles from the deep-water-wave theory are remarkable. However, Fig.~\ref{fig:06}(d) reveals that the maximum values of the horizontal velocity are located below the surface, which is slightly different from the predictions of the inviscid Airy theory. On the contrary, the vertical velocity field is in good agreement with the deep-water velocity profile. Trajectories of SPH particles have also shown deep-water-wave-like behaviour, as depicted in Fig.~\ref{fig:06}(f). Trajectories in the vertical component describe ellipses in good agreement with the deep-water theory. However, trajectories in the horizontal component exhibit slight deviations from the Airy theory.


\subsection{Linearised hydrodynamic equations in the long-wavelength limit \label{subsec:Linearised-hydrodynamic-equation}}

From the numerical simulations of the previous section, we conclude that waves generated by pedal-wavemakers can emulate deep-water waves in a finite-depth tank. However, we observed some deviations from the Airy theory in the horizontal components of the velocity field. This mismatch is related to the effect of viscosity and the formation of boundary layers at the bottom and below the surface, as will be shown.

To solve the linear version of the system \eqref{eq:Ns-incompress}, we use the Helmholtz decomposition $\ensuremath{\mathbf{u}=\boldsymbol{\nabla}\phi+\boldsymbol{\nabla}\times\left(\psi\mathbf{\hat{z}}\right)}$, where $\phi$ is the velocity potential and $\psi$ is the stream function, with $\ensuremath{\mathbf{\hat{z}}\equiv\mathbf{\hat{x}}\times\mathbf{\hat{y}}}$ \cite{Lamb1932}. Thus, the linearized system \eqref{eq:Ns-incompress} decouples $\phi,\psi$ and $P$ in the bulk of the fluid via the three equations:
\begin{subequations}
	\label{Eq:LinearisedEquations}
	\begin{equation}
		\label{eq:incompress2}
		\nabla^2\phi =0,
	\end{equation}
		\begin{equation}
		\label{eq:stream}
		\partial_t\psi-\nu\nabla^2\psi =0,
	\end{equation}
	\begin{equation}
		\label{eq:potential}
		\partial_t\phi+\frac{P}{\rho}+gy=0.
	\end{equation}
\end{subequations}
The linearised boundary condition at the top interface yields
\begin{subequations}
	\label{eq:boundary}
	\begin{equation}
		\label{eq:kinematic}
		\left.\partial_t\eta-\partial_y\phi+\partial_x\psi\right|_{y=\eta}=0,
	\end{equation}
	\begin{equation}
		\label{eq:normal_stress}
		\left.2\nu\left(\partial_{yy}\phi-\partial_{xy}\psi\right)-\frac{P}{\rho}\right|_{y=\eta}=0,
	\end{equation}
	\begin{equation}
		\label{eq:tangential_stress}
		\left.2\partial_{xy}\phi-\partial_{xx}\psi+\partial_{yy}\psi\right|_{y=\eta}=0.
	\end{equation}
\end{subequations}
Equation \eqref{eq:kinematic} is the kinematic condition, whereas equations \eqref{eq:normal_stress} and \eqref{eq:tangential_stress} are the conditions for the normal and tangential stresses, respectively. At the bottom of the wave tank, we assume the following no-slip boundary conditions:
\begin{subequations}
	\label{eq:bottomboundary}
	\begin{equation}
		\label{eq:noslip1}
		\left.\partial_t\chi-\partial_x\phi-\partial_y\psi\right|_{y=-h} =0,
	\end{equation}
		\begin{equation}
		\label{eq:noslip2}
		\left.\partial_t\zeta-\partial_y\phi+\partial_x\psi\right|_{y=-h} =0.
	\end{equation}
\end{subequations}

The system of equations \eqref{Eq:LinearisedEquations} together with the boundary conditions of Eqs. \eqref{eq:boundary} and \eqref{eq:bottomboundary} determines the linear response of the system. The general solutions of the bulk equations \eqref{eq:incompress2} and \eqref{eq:stream} are
\begin{subequations}
	\label{eq:solutions}
	\begin{equation}
		\label{eq:sol_potential}
		\phi(x,\,y,\,t)=\frac{\omega}{k}\Re\left[\left(A\cosh ky+B\sinh ky\right)e^{i(kx-\omega t)}\right],
	\end{equation}
		\begin{equation}
		\label{eq:sol_stream}
		\psi(x,\,y,\,t)=\frac{\omega}{k}\Re\left[\left(C\cosh my+D\sinh my\right)e^{i(kx-\omega t)}\right],
	\end{equation}
\end{subequations}
with $m^{2}=k^{2}-i\omega/\nu$. The values of the complex constants $(A,\,B,\,C,\,D)$ are determined by the boundary conditions. Hereon, it is convenient to introduce the following dimensionless quantities %
\begin{equation}
\bar k\equiv kh,\quad\quad\bar{m}\equiv mh,\quad\quad\bar{\lambda}\equiv \frac{\lambda}{h},\quad\bar\omega\equiv\frac{\omega}{\omega_{\infty}},\label{eq:primes}
\end{equation}
where the characteristic frequency $\omega_{\infty}$ is given by the dispersion relation for deep-water waves, i.e. $\omega_{\infty}^{2}=gk$. Let $\bar\theta\equiv2\nu k^2/\omega$ be the dimensionless viscosity and $\alpha\equiv m/k$ the ratio between the wavenumbers $m$ and $k$. Notice that $\bar\theta$ could also be interpreted as a squared comparison between the extent of an oscillatory Stokes boundary layers, $\sqrt{2\nu/\omega}$, and the wavelength $\lambda$. Evaluating the bulk solutions \eqref{eq:solutions} in the system \eqref{Eq:LinearisedEquations} and normalizing the spatial coordinates according to $\bar x\equiv x/h$ and $\bar y\equiv y/h$, one obtains the following linear system of equations
\begin{equation}
\begin{pmatrix}\bar\omega^{2}\bar\theta\beta & i & 1 & -i\bar\omega^{2}\bar\theta\alpha\\
0 & i & \beta & 0\\
i\cosh\bar k & -i\sinh\bar k & -\alpha\sinh\bar m & \alpha\cosh\bar m\\
-i\sinh\bar k & i\cosh\bar k & \cosh\bar m & -\sinh\bar m
\end{pmatrix}\begin{pmatrix}\bar A\\
\bar B\\
\bar C\\
\bar D
\end{pmatrix}=\begin{pmatrix}0\\
0\\
\bar x_{b}\\
\bar y_{b}
\end{pmatrix},\label{eq:LinearSystem}
\end{equation}
with $\beta\equiv1-i/\bar{\theta}$, $\bar x_b\equiv x_b/h$ and $\bar y_b\equiv y_b/h$. Solving Eq.~(\ref{eq:LinearSystem}) for the dimensionless vector $(\bar A,\,\bar B,\,\bar C,\,\bar D)^{T}\equiv( A,\, B,\,C,\, D)^{T}/h$, one obtains from Eq.~\eqref{eq:solutions} the velocity potential and stream functions,
\begin{subequations}
	\label{eq:FunctionSolutions}
\begin{equation}
    \label{Eq:Potential}
    \begin{split}
        \phi(x,y,t)&=\frac{\omega}{k}\Re\left\{\left[\frac{i\alpha\bar\theta}{\Gamma_B^{(0)}(\bar\theta -i)}\left(\mathcal{F}_p+\frac{\left(\alpha\bar\theta^2\bar\omega^2\Gamma_B^{(1)}-\Gamma_B^{(0)}-\Gamma_S^{(0)}\right)\mathcal{P}}{\alpha^2\bar\theta^2\bar\omega^2(\Gamma_S+\bar\theta\Gamma_B)}\right)\cosh(ky)\right.\right. \\&+\left.\left.\frac{(1+i\bar\theta)\mathcal{P}}{\alpha\bar\theta(\Gamma_S+\bar\theta\Gamma_B)}\sinh(ky)\right] e^{i(kx-\omega t)}\right\},
    \end{split}
\end{equation}
\begin{equation}
    \label{Eq:StreamingFunction}
    \psi(x,y,t)=\frac{\omega}{k}\Re\left[\frac{\mathcal{P} e^{my}}{\alpha(\Gamma_S+\bar\theta\Gamma_B)}e^{i(kx-\omega t)}\right].
\end{equation}
\end{subequations}
The constants $\mathcal{F}_p$ and $\mathcal{P}$ in Eq.~\eqref{eq:FunctionSolutions}, related to the pedal-wavemaking motion at the bottom, are given by
\begin{subequations}
    \begin{equation}
        \mathcal{F}_p\equiv x_b\sinh\bar k+y_b\cosh\bar{k},
    \end{equation}
    \begin{equation}
        \mathcal{P}\equiv\alpha\bar\theta\mathcal{F}_p-(\bar\theta -i)(x_b\sinh\bar m+\alpha y_b\cosh\bar m).
    \end{equation}
\end{subequations}
The constants $\Gamma_S^{(0)}$ and $\Gamma_S$ are related to the angular frequency, depth, wavenumbers and viscosity through,
\begin{subequations}
    \begin{equation}
        \Gamma_S^{(0)}\equiv(i-\bar\theta)\alpha\bar\theta\bar\omega^2,
    \end{equation}
    \begin{equation}
        \Gamma_S\equiv(i-\bar\theta)\left(1+\alpha\bar\theta^2\bar\omega^2\right).
    \end{equation}
\end{subequations}
Finally, the constants $\Gamma_B^{(0)}$, $\Gamma_B^{(1)}$, and $\Gamma_B$ are given by,
\begin{subequations}
\begin{equation}
    \Gamma_B^{(0)}\equiv\alpha\sinh\bar k\cosh\bar m-\cosh\bar k\sinh\bar m,
\end{equation}
\begin{equation}
    \Gamma_B^{(1)}\equiv\alpha\sinh\bar k\sinh\bar m-\cosh\bar k\cosh\bar m,
\end{equation}
\begin{equation}
    \Gamma_B\equiv\Gamma_B^{(0)}-\Gamma_B^{(1)}+\sigma(\alpha\cosh\bar k \cosh\bar m-\sinh\bar k\sinh\bar m),
\end{equation}
\end{subequations}
with $\sigma\equiv\bar\omega^2(\bar\theta-i)^2$.

The velocity fields shown in Figs.~\ref{fig:06}(g) and \ref{fig:06}(h) were obtained from Eqs.~\eqref{eq:FunctionSolutions} and agree well with SPH numerical simulations. Thus, the pedal-wavemaker theory predicts deep-water-like behaviour in the vertical velocity field. Moreover, it also predicts maximum horizontal velocity values below the free surface due to a top boundary layer, showing that the feature observed in SPH simulations is consistent with the effect of viscosity.

\section{Discussion\label{subsec:Discussion}}


The theoretical predictions for the horizontal and vertical components of the velocity of a fluid particle as a function of $\bar y$ for the parameters of the numerical simulation of Section~\ref{subsec:Emulating} are depicted in Fig. \ref{fig:06}(i). Notice that the amplitude in the oscillations of $\bar{u}$ and $\bar{v}$ decays almost linearly with the depth, which is a characteristic
behaviour of shallow-water waves \cite{Lighthill1978}. Indeed, we expect a shallow-water profile given that the wavelength is large compared to the depth of the wave tank. However, the linear decay in the oscillations of $\bar{u}$ and $\bar{v}$ can be regarded as an exponential decay in a large domain truncated to the first-order in depth. Thus, Figs. \ref{fig:06}(c), \ref{fig:06}(f), and \ref{fig:06}(i) can be regarded as the trajectories of fluid particles near the surface of a deep-water wave truncated from below by the wave tank. We conclude that the pedal-wavemaking technique emulates with good accuracy the upper fluid layers of deep-water-waves using a shallow wave tank.

\begin{figure}
\includegraphics[scale=0.56]{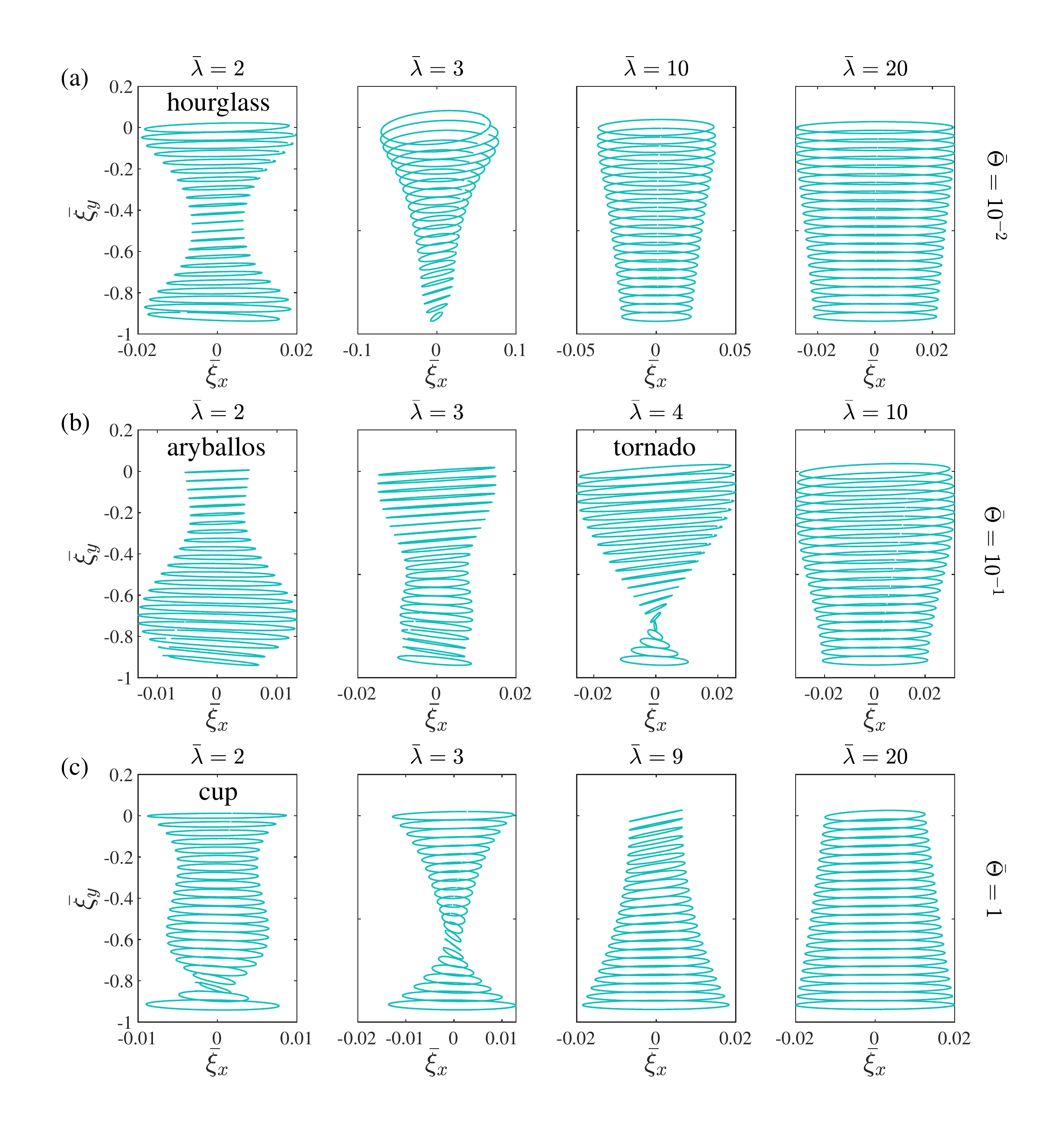}
\caption{Oscillatory particle excursion $\bar\xi_y$ using the pedal-wavemaking technique with $\bar\omega=1$ and $\bar x_b=\bar y_b=1/24$ for (a) $\bar{\Theta}=10^{-2}$, (b) $\bar{\Theta}=10^{-1}$, and (c) $\bar{\Theta}=1$. The normalized wavelength of the generated gravity wave is indicated for each case. \label{fig:07}}
\end{figure}

The form of the velocity potential and streaming functions of Eq.~\eqref{eq:FunctionSolutions} leads us to conclude that the dynamics of fluid particles depends on the viscosity, depth, angular frequency and the wavelength. For practical purposes, we introduce the dimensionless number $\bar \Theta \equiv 2\nu/\sqrt{gh^3}=\bar\theta \bar\omega / \bar k^{3/2}$ which is independent of $\omega$ and $k$. Figure \ref{fig:07} summarises the different type of behaviours obtained through the systematic span of the $\bar{(\lambda},\bar{\Theta)}$-space  (we fixed $\bar \omega = 1$ in all the cases). Figure \ref{fig:07}(a) shows the case of small viscosity, namely $\bar{\Theta}=10^{-2}$. For $\bar{\lambda}=2$, a boundary layer becomes evident at the bottom and below the surface, resulting in an hourglass-like shape in the collective orbits. In this hourglass-like case, the amplitude of the surface wave is nearly the same as the amplitude of the pedalling motion. The behaviour changes dramatically for $\bar{\lambda}=3$, exhibiting an almost deep-water-like behaviour with a significant gain and a thin boundary layer at the bottom. As the wavelength further increases, the system progressively emulates the upper layers of a truncated deep-water behaviour, which is qualitatively similar to the cases shown in Fig. \ref{fig:06}.

Figure \ref{fig:07}(b) shows the particle orbits for moderate viscosity, namely $\bar{\Theta}=10^{-1}$. For $\bar{\lambda}=2$, one obtains a configuration in the form of an Aryballos jar with a thick boundary layer at the bottom of the wave tank. Due to viscosity, the amplitude of the surface wave is smaller than the amplitude of the pedalling motion. The boundary layer at the bottom decreases as the wavelength increases. For $\bar{\lambda}=4$, the gain in the amplitude of the surface wave is large, and the collective configuration of orbits show a rotation of the elliptical axes, exhibiting a tornado-like shape. This feature becomes evident around $\bar{y}=-0.7$, where the slope of the principal axis of the orbits reaches a maximum value and changes its sign. For $\bar{\lambda}=10$, the system emulates the upper layers of a truncated deep-water behaviour with a thin boundary layer at the bottom.

Finally, Fig. \ref{fig:07}(c) shows the particle orbits for large viscosity, namely $\bar{\Theta}=1$. For $\bar{\lambda}=2$, the collective configuration of orbits exhibits a cup-like shape. The slope of the principal axis of the closed orbits reaches a maximum value near the bottom. However, contrary to the tornado-like configuration, there is no change in the orientation of the axes of the ellipses. For $\bar{\lambda}=9$, the system exhibits an almost deep-water-like behaviour with gain smaller than unity due to viscosity. The system emulates deep-water behaviour for larger values of the wavelength.

In summary, from our numerical and mathematical modelling of water waves, we conclude that the pedal-wavemaking technique can emulate deep-water behaviour in a shallow wave tank. Because of viscosity, thin boundary layers appears near the bottom and below the surface, producing a mismatch in the horizontal velocity with inviscid deep-water theory. Notwithstanding, the vertical velocity field using pedal-wavemakers is almost identical to the correspondent velocity field in inviscid deep-water waves.

\section{Conclusions\label{sec:Conclusions}}


We have introduced a new technique to recreate deep long gravity waves in finite-depth water channels using pedal wavemakers at the bottom. The technique consists in moving small segments of the seabed with a  periodic motion, resulting in a multi-point
pedal-like action. The underlying mechanism is efficient in the generation of long wavelengths, which has been elusive using conventional wavemakers \cite{Ursell1960}. The generated wave has no longitudinal loss of wave amplitude due to artificial dissipation in SPH simulations. Studying the linear response of the water free-surface to the pedal like motion in the bottom, we obtained results in full agreement with SPH simulations that emulate deep-water behaviour according to the linear Airy theory. We demonstrate that the problem of evanescent waves appearing near hinged-paddle wavemakers is converted into a thin boundary layer below the surface and at the bottom using pedal-wavemaking. No evanescent waves are present on the surface when the waves are generated by the pedal-wavemakers.

Our wave-generation technique gives useful guidelines to recreate deep-wave phenomena in both experimental and numerical setups with minimum resources. Applications may be found in the study of floating-structures interacting with regular waves under realistic oceanic conditions.

\section*{Acknowledgements}

I.V. and L.G. were funded by Fondecyt/Iniciaci\'on Grant No. 11170700. B.C. was funded by the Australian Research Council Linkage Project Number LP160100391. J.F.M. was funded by Universidad de Santiago de Chile through the POSTDOC\_DICYT Grant Number 042031GZ\_POSTDOC and ANID FONDECYT/POSTDOCTORADO/3200499.

\bibliography{apssamp}

\providecommand{\noopsort}[1]{}\providecommand{\singleletter}[1]{#1}%
\begin{thebibliography}{41}%
\makeatletter
\providecommand \@ifxundefined [1]{%
 \@ifx{#1\undefined}
}%
\providecommand \@ifnum [1]{%
 \ifnum #1\expandafter \@firstoftwo
 \else \expandafter \@secondoftwo
 \fi
}%
\providecommand \@ifx [1]{%
 \ifx #1\expandafter \@firstoftwo
 \else \expandafter \@secondoftwo
 \fi
}%
\providecommand \natexlab [1]{#1}%
\providecommand \enquote  [1]{``#1''}%
\providecommand \bibnamefont  [1]{#1}%
\providecommand \bibfnamefont [1]{#1}%
\providecommand \citenamefont [1]{#1}%
\providecommand \href@noop [0]{\@secondoftwo}%
\providecommand \href [0]{\begingroup \@sanitize@url \@href}%
\providecommand \@href[1]{\@@startlink{#1}\@@href}%
\providecommand \@@href[1]{\endgroup#1\@@endlink}%
\providecommand \@sanitize@url [0]{\catcode `\\12\catcode `\$12\catcode
  `\&12\catcode `\#12\catcode `\^12\catcode `\_12\catcode `\%12\relax}%
\providecommand \@@startlink[1]{}%
\providecommand \@@endlink[0]{}%
\providecommand \url  [0]{\begingroup\@sanitize@url \@url }%
\providecommand \@url [1]{\endgroup\@href {#1}{\urlprefix }}%
\providecommand \urlprefix  [0]{URL }%
\providecommand \Eprint [0]{\href }%
\providecommand \doibase [0]{http://dx.doi.org/}%
\providecommand \selectlanguage [0]{\@gobble}%
\providecommand \bibinfo  [0]{\@secondoftwo}%
\providecommand \bibfield  [0]{\@secondoftwo}%
\providecommand \translation [1]{[#1]}%
\providecommand \BibitemOpen [0]{}%
\providecommand \bibitemStop [0]{}%
\providecommand \bibitemNoStop [0]{.\EOS\space}%
\providecommand \EOS [0]{\spacefactor3000\relax}%
\providecommand \BibitemShut  [1]{\csname bibitem#1\endcsname}%
\let\auto@bib@innerbib\@empty
\bibitem [{\citenamefont {Cross}\ and\ \citenamefont
  {Greenside}(2009)}]{Cross2009}%
  \BibitemOpen
  \bibfield  {author} {\bibinfo {author} {\bibfnamefont {M.}~\bibnamefont
  {Cross}}\ and\ \bibinfo {author} {\bibfnamefont {H.}~\bibnamefont
  {Greenside}},\ }\href@noop {} {\emph {\bibinfo {title} {Pattern formation and
  dynamics in nonequilibrium systems}}}\ (\bibinfo  {publisher} {Cambridge
  University Press},\ \bibinfo {year} {2009})\BibitemShut {NoStop}%
\bibitem [{\citenamefont {Lighthill}(1978)}]{Lighthill1978}%
  \BibitemOpen
  \bibfield  {author} {\bibinfo {author} {\bibfnamefont {J.}~\bibnamefont
  {Lighthill}},\ }\href@noop {} {\emph {\bibinfo {title} {Waves in fluids}}}\
  (\bibinfo  {publisher} {Cambridge university press},\ \bibinfo {year}
  {1978})\BibitemShut {NoStop}%
\bibitem [{\citenamefont {Grilli}\ \emph {et~al.}(2013)\citenamefont {Grilli},
  \citenamefont {Harris}, \citenamefont {Bakhsh}, \citenamefont {Masterlark},
  \citenamefont {Kyriakopoulos}, \citenamefont {Kirby},\ and\ \citenamefont
  {Shi}}]{Grilli2013}%
  \BibitemOpen
  \bibfield  {author} {\bibinfo {author} {\bibfnamefont {S.~T.}\ \bibnamefont
  {Grilli}}, \bibinfo {author} {\bibfnamefont {J.~C.}\ \bibnamefont {Harris}},
  \bibinfo {author} {\bibfnamefont {T.~S.~T.}\ \bibnamefont {Bakhsh}}, \bibinfo
  {author} {\bibfnamefont {T.~L.}\ \bibnamefont {Masterlark}}, \bibinfo
  {author} {\bibfnamefont {C.}~\bibnamefont {Kyriakopoulos}}, \bibinfo {author}
  {\bibfnamefont {J.~T.}\ \bibnamefont {Kirby}}, \ and\ \bibinfo {author}
  {\bibfnamefont {F.}~\bibnamefont {Shi}},\ }\href {\doibase
  10.1007/s00024-012-0528-y} {\bibfield  {journal} {\bibinfo  {journal} {Pure
  and Applied Geophysics}\ }\textbf {\bibinfo {volume} {170}},\ \bibinfo
  {pages} {1333} (\bibinfo {year} {2013})}\BibitemShut {NoStop}%
\bibitem [{\citenamefont {Jamin}\ \emph {et~al.}(2015)\citenamefont {Jamin},
  \citenamefont {Gordillo}, \citenamefont {Ruiz-Chavarr\'ia}, \citenamefont
  {Berhanu},\ and\ \citenamefont {Falcon}}]{Jamin2015}%
  \BibitemOpen
  \bibfield  {author} {\bibinfo {author} {\bibfnamefont {T.}~\bibnamefont
  {Jamin}}, \bibinfo {author} {\bibfnamefont {L.}~\bibnamefont {Gordillo}},
  \bibinfo {author} {\bibfnamefont {G.}~\bibnamefont {Ruiz-Chavarr\'ia}},
  \bibinfo {author} {\bibfnamefont {M.}~\bibnamefont {Berhanu}}, \ and\
  \bibinfo {author} {\bibfnamefont {E.}~\bibnamefont {Falcon}},\ }\href
  {\doibase 10.1098/rspa.2015.0069} {\bibfield  {journal} {\bibinfo  {journal}
  {Proceedings of the Royal Society A: Mathematical, Physical and Engineering
  Sciences}\ }\textbf {\bibinfo {volume} {471}},\ \bibinfo {pages} {20150069}
  (\bibinfo {year} {2015})}\BibitemShut {NoStop}%
\bibitem [{\citenamefont {Cunningham}\ \emph {et~al.}(2014)\citenamefont
  {Cunningham}, \citenamefont {Rogers},\ and\ \citenamefont
  {Pringgana}}]{Cunningham2014}%
  \BibitemOpen
  \bibfield  {author} {\bibinfo {author} {\bibfnamefont {L.~S.}\ \bibnamefont
  {Cunningham}}, \bibinfo {author} {\bibfnamefont {B.~D.}\ \bibnamefont
  {Rogers}}, \ and\ \bibinfo {author} {\bibfnamefont {G.}~\bibnamefont
  {Pringgana}},\ }\href {\doibase 10.1680/eacm.13.00028} {\bibfield  {journal}
  {\bibinfo  {journal} {Proceedings of the Institution of Civil Engineers -
  Engineering and Computational Mechanics}\ }\textbf {\bibinfo {volume}
  {167}},\ \bibinfo {pages} {126} (\bibinfo {year} {2014})}\BibitemShut
  {NoStop}%
\bibitem [{\citenamefont {Groenenboom}\ and\ \citenamefont
  {Cartwright}(2010)}]{Cartwright2010}%
  \BibitemOpen
  \bibfield  {author} {\bibinfo {author} {\bibfnamefont {P.~H.}\ \bibnamefont
  {Groenenboom}}\ and\ \bibinfo {author} {\bibfnamefont {B.~K.}\ \bibnamefont
  {Cartwright}},\ }\href {\doibase 10.1080/00221686.2010.9641246} {\bibfield
  {journal} {\bibinfo  {journal} {Journal of Hydraulic Research}\ }\textbf
  {\bibinfo {volume} {48}},\ \bibinfo {pages} {61} (\bibinfo {year}
  {2010})}\BibitemShut {NoStop}%
\bibitem [{\citenamefont {Cartwright}(2012)}]{Cartwright2012}%
  \BibitemOpen
  \bibfield  {author} {\bibinfo {author} {\bibfnamefont {B.~K.}\ \bibnamefont
  {Cartwright}},\ }\emph {\bibinfo {title} {The study of ship motions in
  regular waves using a mesh-free numerical method}},\ \href@noop {} {Master's
  thesis},\ \bibinfo  {school} {University of Tasmania} (\bibinfo {year}
  {2012})\BibitemShut {NoStop}%
\bibitem [{\citenamefont {Lloyd}\ \emph {et~al.}(1991)\citenamefont {Lloyd},
  \citenamefont {Hosoda}, \citenamefont {Robinson}, \citenamefont {Nicholson},
  \citenamefont {Victory}, \citenamefont {Price}, \citenamefont {Bishop},\ and\
  \citenamefont {Price}}]{Lloyd1991}%
  \BibitemOpen
  \bibfield  {author} {\bibinfo {author} {\bibfnamefont {A.~R. J.~M.}\
  \bibnamefont {Lloyd}}, \bibinfo {author} {\bibfnamefont {R.}~\bibnamefont
  {Hosoda}}, \bibinfo {author} {\bibfnamefont {D.~W.}\ \bibnamefont
  {Robinson}}, \bibinfo {author} {\bibfnamefont {K.}~\bibnamefont {Nicholson}},
  \bibinfo {author} {\bibfnamefont {G.}~\bibnamefont {Victory}}, \bibinfo
  {author} {\bibfnamefont {W.~G.}\ \bibnamefont {Price}}, \bibinfo {author}
  {\bibfnamefont {R.~E.~D.}\ \bibnamefont {Bishop}}, \ and\ \bibinfo {author}
  {\bibfnamefont {W.~G.}\ \bibnamefont {Price}},\ }\href {\doibase
  10.1098/rsta.1991.0012} {\bibfield  {journal} {\bibinfo  {journal}
  {Philosophical Transactions of the Royal Society of London. Series A:
  Physical and Engineering Sciences}\ }\textbf {\bibinfo {volume} {334}},\
  \bibinfo {pages} {253} (\bibinfo {year} {1991})}\BibitemShut {NoStop}%
\bibitem [{\citenamefont {Li}\ and\ \citenamefont {Liu}(2002)}]{Li2002}%
  \BibitemOpen
  \bibfield  {author} {\bibinfo {author} {\bibfnamefont {S.}~\bibnamefont
  {Li}}\ and\ \bibinfo {author} {\bibfnamefont {W.~K.}\ \bibnamefont {Liu}},\
  }\href {\doibase doi:10.1115/1.1431547} {\bibfield  {journal} {\bibinfo
  {journal} {Applied Mechanics Reviews}\ }\textbf {\bibinfo {volume} {55}},\
  \bibinfo {pages} {1} (\bibinfo {year} {2002})}\BibitemShut {NoStop}%
\bibitem [{\citenamefont {Liu}\ and\ \citenamefont {Liu}(2003)}]{Liu2003}%
  \BibitemOpen
  \bibfield  {author} {\bibinfo {author} {\bibfnamefont {G.-R.}\ \bibnamefont
  {Liu}}\ and\ \bibinfo {author} {\bibfnamefont {M.~B.}\ \bibnamefont {Liu}},\
  }\href@noop {} {\emph {\bibinfo {title} {Smoothed particle hydrodynamics: a
  meshfree particle method}}}\ (\bibinfo  {publisher} {World scientific},\
  \bibinfo {year} {2003})\BibitemShut {NoStop}%
\bibitem [{\citenamefont {Sigalotti}\ \emph {et~al.}(2003)\citenamefont
  {Sigalotti}, \citenamefont {Klapp}, \citenamefont {Sira}, \citenamefont
  {Mele\'an},\ and\ \citenamefont {Hasmy}}]{Sigalotti2003}%
  \BibitemOpen
  \bibfield  {author} {\bibinfo {author} {\bibfnamefont {L.~D.~G.}\
  \bibnamefont {Sigalotti}}, \bibinfo {author} {\bibfnamefont {J.}~\bibnamefont
  {Klapp}}, \bibinfo {author} {\bibfnamefont {E.}~\bibnamefont {Sira}},
  \bibinfo {author} {\bibfnamefont {Y.}~\bibnamefont {Mele\'an}}, \ and\
  \bibinfo {author} {\bibfnamefont {A.}~\bibnamefont {Hasmy}},\ }\href
  {\doibase https://doi.org/10.1016/S0021-9991(03)00343-7} {\bibfield
  {journal} {\bibinfo  {journal} {Journal of Computational Physics}\ }\textbf
  {\bibinfo {volume} {191}},\ \bibinfo {pages} {622 } (\bibinfo {year}
  {2003})}\BibitemShut {NoStop}%
\bibitem [{\citenamefont {Liu}\ and\ \citenamefont {Liu}(2010)}]{Liu2010}%
  \BibitemOpen
  \bibfield  {author} {\bibinfo {author} {\bibfnamefont {M.~B.}\ \bibnamefont
  {Liu}}\ and\ \bibinfo {author} {\bibfnamefont {G.~R.}\ \bibnamefont {Liu}},\
  }\href {\doibase 10.1007/s11831-010-9040-7} {\bibfield  {journal} {\bibinfo
  {journal} {Archives of Computational Methods in Engineering}\ }\textbf
  {\bibinfo {volume} {17}},\ \bibinfo {pages} {25} (\bibinfo {year}
  {2010})}\BibitemShut {NoStop}%
\bibitem [{\citenamefont {Wang}\ \emph {et~al.}(2016)\citenamefont {Wang},
  \citenamefont {Chen}, \citenamefont {Wang}, \citenamefont {Liao},
  \citenamefont {Zhu},\ and\ \citenamefont {Li}}]{Wang2016}%
  \BibitemOpen
  \bibfield  {author} {\bibinfo {author} {\bibfnamefont {Z.-B.}\ \bibnamefont
  {Wang}}, \bibinfo {author} {\bibfnamefont {R.}~\bibnamefont {Chen}}, \bibinfo
  {author} {\bibfnamefont {H.}~\bibnamefont {Wang}}, \bibinfo {author}
  {\bibfnamefont {Q.}~\bibnamefont {Liao}}, \bibinfo {author} {\bibfnamefont
  {X.}~\bibnamefont {Zhu}}, \ and\ \bibinfo {author} {\bibfnamefont {S.-Z.}\
  \bibnamefont {Li}},\ }\href {\doibase
  https://doi.org/10.1016/j.apm.2016.06.030} {\bibfield  {journal} {\bibinfo
  {journal} {Applied Mathematical Modelling}\ }\textbf {\bibinfo {volume}
  {40}},\ \bibinfo {pages} {9625 } (\bibinfo {year} {2016})}\BibitemShut
  {NoStop}%
\bibitem [{\citenamefont {Sigalotti}\ \emph {et~al.}(2009)\citenamefont
  {Sigalotti}, \citenamefont {L\'opez},\ and\ \citenamefont
  {Trujillo}}]{Sigalotti2009}%
  \BibitemOpen
  \bibfield  {author} {\bibinfo {author} {\bibfnamefont {L.~D.~G.}\
  \bibnamefont {Sigalotti}}, \bibinfo {author} {\bibfnamefont {H.}~\bibnamefont
  {L\'opez}}, \ and\ \bibinfo {author} {\bibfnamefont {L.}~\bibnamefont
  {Trujillo}},\ }\href {\doibase https://doi.org/10.1016/j.jcp.2009.04.041}
  {\bibfield  {journal} {\bibinfo  {journal} {Journal of Computational
  Physics}\ }\textbf {\bibinfo {volume} {228}},\ \bibinfo {pages} {5888 }
  (\bibinfo {year} {2009})}\BibitemShut {NoStop}%
\bibitem [{\citenamefont {Sigalotti}\ \emph {et~al.}(2006)\citenamefont
  {Sigalotti}, \citenamefont {L\'opez}, \citenamefont {Donoso}, \citenamefont
  {Sira},\ and\ \citenamefont {Klapp}}]{Sigalotti2006}%
  \BibitemOpen
  \bibfield  {author} {\bibinfo {author} {\bibfnamefont {L.~D.~G.}\
  \bibnamefont {Sigalotti}}, \bibinfo {author} {\bibfnamefont {H.}~\bibnamefont
  {L\'opez}}, \bibinfo {author} {\bibfnamefont {A.}~\bibnamefont {Donoso}},
  \bibinfo {author} {\bibfnamefont {E.}~\bibnamefont {Sira}}, \ and\ \bibinfo
  {author} {\bibfnamefont {J.}~\bibnamefont {Klapp}},\ }\href {\doibase
  https://doi.org/10.1016/j.jcp.2005.06.016} {\bibfield  {journal} {\bibinfo
  {journal} {Journal of Computational Physics}\ }\textbf {\bibinfo {volume}
  {212}},\ \bibinfo {pages} {124 } (\bibinfo {year} {2006})}\BibitemShut
  {NoStop}%
\bibitem [{\citenamefont {Marrone}\ \emph {et~al.}(2011)\citenamefont
  {Marrone}, \citenamefont {Antuono}, \citenamefont {Colagrossi}, \citenamefont
  {Colicchio}, \citenamefont {Touz\'e},\ and\ \citenamefont
  {Graziani}}]{Marrone2011}%
  \BibitemOpen
  \bibfield  {author} {\bibinfo {author} {\bibfnamefont {S.}~\bibnamefont
  {Marrone}}, \bibinfo {author} {\bibfnamefont {M.}~\bibnamefont {Antuono}},
  \bibinfo {author} {\bibfnamefont {A.}~\bibnamefont {Colagrossi}}, \bibinfo
  {author} {\bibfnamefont {G.}~\bibnamefont {Colicchio}}, \bibinfo {author}
  {\bibfnamefont {D.~L.}\ \bibnamefont {Touz\'e}}, \ and\ \bibinfo {author}
  {\bibfnamefont {G.}~\bibnamefont {Graziani}},\ }\href {\doibase
  https://doi.org/10.1016/j.cma.2010.12.016} {\bibfield  {journal} {\bibinfo
  {journal} {Computer Methods in Applied Mechanics and Engineering}\ }\textbf
  {\bibinfo {volume} {200}},\ \bibinfo {pages} {1526} (\bibinfo {year}
  {2011})}\BibitemShut {NoStop}%
\bibitem [{\citenamefont {Sun}\ \emph {et~al.}(2018)\citenamefont {Sun},
  \citenamefont {Colagrossi}, \citenamefont {Marrone}, \citenamefont
  {Antuono},\ and\ \citenamefont {Zhang}}]{Sun2018}%
  \BibitemOpen
  \bibfield  {author} {\bibinfo {author} {\bibfnamefont {P.}~\bibnamefont
  {Sun}}, \bibinfo {author} {\bibfnamefont {A.}~\bibnamefont {Colagrossi}},
  \bibinfo {author} {\bibfnamefont {S.}~\bibnamefont {Marrone}}, \bibinfo
  {author} {\bibfnamefont {M.}~\bibnamefont {Antuono}}, \ and\ \bibinfo
  {author} {\bibfnamefont {A.}~\bibnamefont {Zhang}},\ }\href {\doibase
  https://doi.org/10.1016/j.cpc.2017.11.016} {\bibfield  {journal} {\bibinfo
  {journal} {Computer Physics Communications}\ }\textbf {\bibinfo {volume}
  {224}},\ \bibinfo {pages} {63 } (\bibinfo {year} {2018})}\BibitemShut
  {NoStop}%
\bibitem [{\citenamefont {De~Padova}\ \emph {et~al.}(2020)\citenamefont
  {De~Padova}, \citenamefont {Mossa},\ and\ \citenamefont
  {Sibilla}}]{dePadova2020}%
  \BibitemOpen
  \bibfield  {author} {\bibinfo {author} {\bibfnamefont {D.}~\bibnamefont
  {De~Padova}}, \bibinfo {author} {\bibfnamefont {M.}~\bibnamefont {Mossa}}, \
  and\ \bibinfo {author} {\bibfnamefont {S.}~\bibnamefont {Sibilla}},\
  }\href@noop {} {\bibfield  {journal} {\bibinfo  {journal} {Environmental
  Fluid Mechanics}\ }\textbf {\bibinfo {volume} {20}},\ \bibinfo {pages} {189}
  (\bibinfo {year} {2020})}\BibitemShut {NoStop}%
\bibitem [{\citenamefont {Mele\'an}\ \emph {et~al.}(2004)\citenamefont
  {Mele\'an}, \citenamefont {Sigalotti},\ and\ \citenamefont
  {Hasmy}}]{Melean2004}%
  \BibitemOpen
  \bibfield  {author} {\bibinfo {author} {\bibfnamefont {Y.}~\bibnamefont
  {Mele\'an}}, \bibinfo {author} {\bibfnamefont {L.~D.~G.}\ \bibnamefont
  {Sigalotti}}, \ and\ \bibinfo {author} {\bibfnamefont {A.}~\bibnamefont
  {Hasmy}},\ }\href {\doibase https://doi.org/10.1016/j.comphy.2003.11.002}
  {\bibfield  {journal} {\bibinfo  {journal} {Computer Physics Communications}\
  }\textbf {\bibinfo {volume} {157}},\ \bibinfo {pages} {191 } (\bibinfo {year}
  {2004})}\BibitemShut {NoStop}%
\bibitem [{\citenamefont {Meleán}\ and\ \citenamefont
  {Sigalotti}(2005)}]{Melean2005}%
  \BibitemOpen
  \bibfield  {author} {\bibinfo {author} {\bibfnamefont {Y.}~\bibnamefont
  {Meleán}}\ and\ \bibinfo {author} {\bibfnamefont {L.~D.~G.}\ \bibnamefont
  {Sigalotti}},\ }\href {\doibase
  https://doi.org/10.1016/j.ijheatmasstransfer.2005.04.006} {\bibfield
  {journal} {\bibinfo  {journal} {International Journal of Heat and Mass
  Transfer}\ }\textbf {\bibinfo {volume} {48}},\ \bibinfo {pages} {4041 }
  (\bibinfo {year} {2005})}\BibitemShut {NoStop}%
\bibitem [{\citenamefont {Ming}\ \emph {et~al.}(2017)\citenamefont {Ming},
  \citenamefont {Sun},\ and\ \citenamefont {Zhang}}]{Ming2017}%
  \BibitemOpen
  \bibfield  {author} {\bibinfo {author} {\bibfnamefont {F.}~\bibnamefont
  {Ming}}, \bibinfo {author} {\bibfnamefont {P.}~\bibnamefont {Sun}}, \ and\
  \bibinfo {author} {\bibfnamefont {A.}~\bibnamefont {Zhang}},\ }\href@noop {}
  {\bibfield  {journal} {\bibinfo  {journal} {Meccanica}\ }\textbf {\bibinfo
  {volume} {52}},\ \bibinfo {pages} {2665} (\bibinfo {year}
  {2017})}\BibitemShut {NoStop}%
\bibitem [{\citenamefont {Gnanasekaran}\ \emph {et~al.}(2019)\citenamefont
  {Gnanasekaran}, \citenamefont {Liu}, \citenamefont {Fu}, \citenamefont
  {Wang}, \citenamefont {Niu},\ and\ \citenamefont {Lin}}]{Gnanasekaran2019}%
  \BibitemOpen
  \bibfield  {author} {\bibinfo {author} {\bibfnamefont {B.}~\bibnamefont
  {Gnanasekaran}}, \bibinfo {author} {\bibfnamefont {G.-R.}\ \bibnamefont
  {Liu}}, \bibinfo {author} {\bibfnamefont {Y.}~\bibnamefont {Fu}}, \bibinfo
  {author} {\bibfnamefont {G.}~\bibnamefont {Wang}}, \bibinfo {author}
  {\bibfnamefont {W.}~\bibnamefont {Niu}}, \ and\ \bibinfo {author}
  {\bibfnamefont {T.}~\bibnamefont {Lin}},\ }\href {\doibase
  https://doi.org/10.1016/j.surfcoat.2019.07.036} {\bibfield  {journal}
  {\bibinfo  {journal} {Surface and Coatings Technology}\ }\textbf {\bibinfo
  {volume} {377}},\ \bibinfo {pages} {124812} (\bibinfo {year}
  {2019})}\BibitemShut {NoStop}%
\bibitem [{\citenamefont {Sigalotti}\ and\ \citenamefont
  {L\'opez}(2008)}]{Sigalotti2008}%
  \BibitemOpen
  \bibfield  {author} {\bibinfo {author} {\bibfnamefont {L.~D.~G.}\
  \bibnamefont {Sigalotti}}\ and\ \bibinfo {author} {\bibfnamefont
  {H.}~\bibnamefont {L\'opez}},\ }\href {\doibase
  https://doi.org/10.1016/j.camwa.2007.03.007} {\bibfield  {journal} {\bibinfo
  {journal} {Computers \& Mathematics with Applications}\ }\textbf {\bibinfo
  {volume} {55}},\ \bibinfo {pages} {23 } (\bibinfo {year} {2008})}\BibitemShut
  {NoStop}%
\bibitem [{\citenamefont {Milgram}(1970)}]{Milgram1970}%
  \BibitemOpen
  \bibfield  {author} {\bibinfo {author} {\bibfnamefont {J.~H.}\ \bibnamefont
  {Milgram}},\ }\href {\doibase 10.1017/S0022112070001635} {\bibfield
  {journal} {\bibinfo  {journal} {J. Fluid Mech.}\ }\textbf {\bibinfo {volume}
  {42}},\ \bibinfo {pages} {845} (\bibinfo {year} {1970})}\BibitemShut
  {NoStop}%
\bibitem [{\citenamefont {Sch\"affer}\ and\ \citenamefont
  {Klopman}(2000)}]{Schaffer2000}%
  \BibitemOpen
  \bibfield  {author} {\bibinfo {author} {\bibfnamefont {H.~A.}\ \bibnamefont
  {Sch\"affer}}\ and\ \bibinfo {author} {\bibfnamefont {G.}~\bibnamefont
  {Klopman}},\ }\href {\doibase 10.1061/(ASCE)0733-950X(2000)126:2(88)}
  {\bibfield  {journal} {\bibinfo  {journal} {J. Waterw. Port C.-Asce}\
  }\textbf {\bibinfo {volume} {126}},\ \bibinfo {pages} {88} (\bibinfo {year}
  {2000})}\BibitemShut {NoStop}%
\bibitem [{\citenamefont {Ouellet}\ and\ \citenamefont
  {Datta}(1986)}]{Ouellet1986}%
  \BibitemOpen
  \bibfield  {author} {\bibinfo {author} {\bibfnamefont {Y.}~\bibnamefont
  {Ouellet}}\ and\ \bibinfo {author} {\bibfnamefont {I.}~\bibnamefont
  {Datta}},\ }\href {\doibase 10.1080/00221688609499305} {\bibfield  {journal}
  {\bibinfo  {journal} {J. Hydraul. Res.}\ }\textbf {\bibinfo {volume} {24}},\
  \bibinfo {pages} {265} (\bibinfo {year} {1986})}\BibitemShut {NoStop}%
\bibitem [{\citenamefont {Havelock}(1929)}]{Havelock1929}%
  \BibitemOpen
  \bibfield  {author} {\bibinfo {author} {\bibfnamefont {T.}~\bibnamefont
  {Havelock}},\ }\href {\doibase 10.1080/14786441008564913} {\bibfield
  {journal} {\bibinfo  {journal} {Lond. Edinb. Dubl. Phil. Mag.}\ }\textbf
  {\bibinfo {volume} {8}},\ \bibinfo {pages} {569} (\bibinfo {year}
  {1929})}\BibitemShut {NoStop}%
\bibitem [{\citenamefont {{Ba Thuy}}\ \emph {et~al.}(2009)\citenamefont {{Ba
  Thuy}}, \citenamefont {Tanimoto}, \citenamefont {Tanaka}, \citenamefont
  {Harada},\ and\ \citenamefont {Iimura}}]{Bathuy2009}%
  \BibitemOpen
  \bibfield  {author} {\bibinfo {author} {\bibfnamefont {N.}~\bibnamefont {{Ba
  Thuy}}}, \bibinfo {author} {\bibfnamefont {K.}~\bibnamefont {Tanimoto}},
  \bibinfo {author} {\bibfnamefont {N.}~\bibnamefont {Tanaka}}, \bibinfo
  {author} {\bibfnamefont {K.}~\bibnamefont {Harada}}, \ and\ \bibinfo {author}
  {\bibfnamefont {K.}~\bibnamefont {Iimura}},\ }\href {\doibase
  https://doi.org/10.1016/j.oceaneng.2009.07.006} {\bibfield  {journal}
  {\bibinfo  {journal} {Ocean Engineering}\ }\textbf {\bibinfo {volume} {36}},\
  \bibinfo {pages} {1258 } (\bibinfo {year} {2009})}\BibitemShut {NoStop}%
\bibitem [{\citenamefont {Briggs}\ \emph {et~al.}(1995)\citenamefont {Briggs},
  \citenamefont {Synolakis}, \citenamefont {Harkins},\ and\ \citenamefont
  {Green}}]{Briggs1995}%
  \BibitemOpen
  \bibfield  {author} {\bibinfo {author} {\bibfnamefont {M.~J.}\ \bibnamefont
  {Briggs}}, \bibinfo {author} {\bibfnamefont {C.~E.}\ \bibnamefont
  {Synolakis}}, \bibinfo {author} {\bibfnamefont {G.~S.}\ \bibnamefont
  {Harkins}}, \ and\ \bibinfo {author} {\bibfnamefont {D.~R.}\ \bibnamefont
  {Green}},\ }\href@noop {} {\bibfield  {journal} {\bibinfo  {journal} {Pure
  and applied geophysics}\ }\textbf {\bibinfo {volume} {144}},\ \bibinfo
  {pages} {569} (\bibinfo {year} {1995})}\BibitemShut {NoStop}%
\bibitem [{\citenamefont {Dematteis}\ \emph {et~al.}(2019)\citenamefont
  {Dematteis}, \citenamefont {Grafke}, \citenamefont {Onorato},\ and\
  \citenamefont {Vanden-Eijnden}}]{Dematteis2019}%
  \BibitemOpen
  \bibfield  {author} {\bibinfo {author} {\bibfnamefont {G.}~\bibnamefont
  {Dematteis}}, \bibinfo {author} {\bibfnamefont {T.}~\bibnamefont {Grafke}},
  \bibinfo {author} {\bibfnamefont {M.}~\bibnamefont {Onorato}}, \ and\
  \bibinfo {author} {\bibfnamefont {E.}~\bibnamefont {Vanden-Eijnden}},\ }\href
  {\doibase 10.1103/PhysRevX.9.041057} {\bibfield  {journal} {\bibinfo
  {journal} {Phys. Rev. X}\ }\textbf {\bibinfo {volume} {9}},\ \bibinfo {pages}
  {041057} (\bibinfo {year} {2019})}\BibitemShut {NoStop}%
\bibitem [{\citenamefont {McAllister}\ \emph {et~al.}(2019)\citenamefont
  {McAllister}, \citenamefont {Draycott}, \citenamefont {Adcock}, \citenamefont
  {Taylor},\ and\ \citenamefont {van~den Bremer}}]{Mcallister2019}%
  \BibitemOpen
  \bibfield  {author} {\bibinfo {author} {\bibfnamefont {M.~L.}\ \bibnamefont
  {McAllister}}, \bibinfo {author} {\bibfnamefont {S.}~\bibnamefont
  {Draycott}}, \bibinfo {author} {\bibfnamefont {T.~A.~A.}\ \bibnamefont
  {Adcock}}, \bibinfo {author} {\bibfnamefont {P.~H.}\ \bibnamefont {Taylor}},
  \ and\ \bibinfo {author} {\bibfnamefont {T.~S.}\ \bibnamefont {van~den
  Bremer}},\ }\href {\doibase 10.1017/jfm.2018.886} {\bibfield  {journal}
  {\bibinfo  {journal} {Journal of Fluid Mechanics}\ }\textbf {\bibinfo
  {volume} {860}},\ \bibinfo {pages} {767} (\bibinfo {year}
  {2019})}\BibitemShut {NoStop}%
\bibitem [{\citenamefont {Ursell}\ \emph {et~al.}(1960)\citenamefont {Ursell},
  \citenamefont {Dean},\ and\ \citenamefont {Yu}}]{Ursell1960}%
  \BibitemOpen
  \bibfield  {author} {\bibinfo {author} {\bibfnamefont {F.}~\bibnamefont
  {Ursell}}, \bibinfo {author} {\bibfnamefont {R.~G.}\ \bibnamefont {Dean}}, \
  and\ \bibinfo {author} {\bibfnamefont {Y.~S.}\ \bibnamefont {Yu}},\ }\href
  {\doibase 10.1017/S0022112060000037} {\bibfield  {journal} {\bibinfo
  {journal} {J. Fluid Mech.}\ }\textbf {\bibinfo {volume} {7}},\ \bibinfo
  {pages} {33} (\bibinfo {year} {1960})}\BibitemShut {NoStop}%
\bibitem [{\citenamefont {Ozbulut}\ \emph {et~al.}(2020)\citenamefont
  {Ozbulut}, \citenamefont {Ramezanzadeh}, \citenamefont {Yildiz},\ and\
  \citenamefont {Goren}}]{Ozbulut2020}%
  \BibitemOpen
  \bibfield  {author} {\bibinfo {author} {\bibfnamefont {M.}~\bibnamefont
  {Ozbulut}}, \bibinfo {author} {\bibfnamefont {S.}~\bibnamefont
  {Ramezanzadeh}}, \bibinfo {author} {\bibfnamefont {M.}~\bibnamefont
  {Yildiz}}, \ and\ \bibinfo {author} {\bibfnamefont {O.}~\bibnamefont
  {Goren}},\ }\href {\doibase doi.org/10.1007/s40722-020-00163-x} {\bibfield
  {journal} {\bibinfo  {journal} {Journal of Ocean Engineering and Marine
  Energy}\ ,\ \bibinfo {pages} {1}} (\bibinfo {year} {2020})}\BibitemShut
  {NoStop}%
\bibitem [{\citenamefont {Stoker}(2011)}]{Stoker2011}%
  \BibitemOpen
  \bibfield  {author} {\bibinfo {author} {\bibfnamefont {J.~J.}\ \bibnamefont
  {Stoker}},\ }\href@noop {} {\emph {\bibinfo {title} {Water waves: The
  mathematical theory with applications}}},\ Vol.~\bibinfo {volume} {36}\
  (\bibinfo  {publisher} {John Wiley \& Sons},\ \bibinfo {year}
  {2011})\BibitemShut {NoStop}%
\bibitem [{\citenamefont {Lamb}(1932)}]{Lamb1932}%
  \BibitemOpen
  \bibfield  {author} {\bibinfo {author} {\bibfnamefont {H.~L.}\ \bibnamefont
  {Lamb}},\ }\href@noop {} {\emph {\bibinfo {title} {Hydrodynamics}}}\
  (\bibinfo  {publisher} {Cambridge University Press},\ \bibinfo {year}
  {1932})\BibitemShut {NoStop}%
\bibitem [{\citenamefont {Kajiura}(1963)}]{Kajiura1963}%
  \BibitemOpen
  \bibfield  {author} {\bibinfo {author} {\bibfnamefont {K.}~\bibnamefont
  {Kajiura}},\ }\href@noop {} {\bibfield  {journal} {\bibinfo  {journal} {B.
  Earthq. Res. I. Tokyo}\ }\textbf {\bibinfo {volume} {41}},\ \bibinfo {pages}
  {535} (\bibinfo {year} {1963})}\BibitemShut {NoStop}%
\bibitem [{\citenamefont {Munk}(1950)}]{Munk1950}%
  \BibitemOpen
  \bibfield  {author} {\bibinfo {author} {\bibfnamefont {W.}~\bibnamefont
  {Munk}},\ }\href {\doibase 10.9753/icce.v1.1} {\bibfield  {journal} {\bibinfo
   {journal} {Coast. Eng. Proc.}\ }\textbf {\bibinfo {volume} {1}},\ \bibinfo
  {pages} {1} (\bibinfo {year} {1950})}\BibitemShut {NoStop}%
\bibitem [{\citenamefont {McCue}\ \emph {et~al.}(2006)\citenamefont {McCue},
  \citenamefont {Alford}, \citenamefont {Belknap}, \citenamefont {Bulian},
  \citenamefont {Delorme}, \citenamefont {Francescutto},\ and\ \citenamefont
  {Vakakis}}]{McCue2006}%
  \BibitemOpen
  \bibfield  {author} {\bibinfo {author} {\bibfnamefont {L.}~\bibnamefont
  {McCue}}, \bibinfo {author} {\bibfnamefont {L.}~\bibnamefont {Alford}},
  \bibinfo {author} {\bibfnamefont {W.}~\bibnamefont {Belknap}}, \bibinfo
  {author} {\bibfnamefont {G.}~\bibnamefont {Bulian}}, \bibinfo {author}
  {\bibfnamefont {L.}~\bibnamefont {Delorme}}, \bibinfo {author} {\bibfnamefont
  {A.}~\bibnamefont {Francescutto}}, \ and\ \bibinfo {author} {\bibfnamefont
  {A.}~\bibnamefont {Vakakis}},\ }\href {\doibase
  https://onepetro.org/journal-paper/SNAME-MTSN-2006-43-1-55} {\bibfield
  {journal} {\bibinfo  {journal} {Proceedings of the 2005 SIAM Conference on
  Applications of Dynamical Systems}\ }\textbf {\bibinfo {volume} {43}},\
  \bibinfo {pages} {55} (\bibinfo {year} {2006})}\BibitemShut {NoStop}%
\bibitem [{\citenamefont {Jones}\ and\ \citenamefont
  {Belton}(2006)}]{Jones2006}%
  \BibitemOpen
  \bibfield  {author} {\bibinfo {author} {\bibfnamefont {D.~A.}\ \bibnamefont
  {Jones}}\ and\ \bibinfo {author} {\bibfnamefont {D.}~\bibnamefont {Belton}},\
  }\href
  {http://dspace.dsto.defence.gov.au/dspace/bitstream/1947/4543/4/DSTO-TR-1922.PR.pdf}
  {\emph {\bibinfo {title} {Smoothed Particle Hydrodynamics: Applications
  within DSTO}}},\ \bibinfo {type} {Tech. Rep.}\ (\bibinfo  {institution}
  {Defence Science Technology Organisation, Australia,},\ \bibinfo {year}
  {2006})\BibitemShut {NoStop}%
\bibitem [{\citenamefont {Guilcher}\ \emph {et~al.}(2007)\citenamefont
  {Guilcher}, \citenamefont {Ducorzet}, \citenamefont {Alessandrini},\ and\
  \citenamefont {Ferrant}}]{Guilcher2007}%
  \BibitemOpen
  \bibfield  {author} {\bibinfo {author} {\bibfnamefont {P.}~\bibnamefont
  {Guilcher}}, \bibinfo {author} {\bibfnamefont {G.}~\bibnamefont {Ducorzet}},
  \bibinfo {author} {\bibfnamefont {B.}~\bibnamefont {Alessandrini}}, \ and\
  \bibinfo {author} {\bibfnamefont {P.}~\bibnamefont {Ferrant}},\ }in\
  \href@noop {} {\emph {\bibinfo {booktitle} {Proceedings of 2nd International
  SPHERIC Workshop}}}\ (\bibinfo {year} {2007})\ pp.\ \bibinfo {pages}
  {119--124}\BibitemShut {NoStop}%
\bibitem [{\citenamefont {De~Padova}\ \emph {et~al.}(2009)\citenamefont
  {De~Padova}, \citenamefont {Dalrymple}, \citenamefont {Mossa},\ and\
  \citenamefont {Petrillo}}]{DePadova2009}%
  \BibitemOpen
  \bibfield  {author} {\bibinfo {author} {\bibfnamefont {D.}~\bibnamefont
  {De~Padova}}, \bibinfo {author} {\bibfnamefont {R.}~\bibnamefont
  {Dalrymple}}, \bibinfo {author} {\bibfnamefont {M.}~\bibnamefont {Mossa}}, \
  and\ \bibinfo {author} {\bibfnamefont {A.}~\bibnamefont {Petrillo}},\
  }\href@noop {} {\bibfield  {journal} {\bibinfo  {journal} {arXiv preprint
  arXiv:0911.1872}\ } (\bibinfo {year} {2009})}\BibitemShut {NoStop}%
\end{thebibliography}%

\end{document}